\documentclass[11pt,letterpaper,usenames,dvipsnames]{article}
\usepackage{jheppub}

\allowdisplaybreaks[2]  
\usepackage{pifont}
\usepackage{bm} 
\usepackage{dsfont} 
\usepackage{float}
\usepackage{hhline}
\usepackage{mathtools}

\usepackage{verbatim}
\usepackage{tikz}
\usetikzlibrary{arrows,snakes,shapes.arrows,decorations.markings}
     \tikzset{>=triangle 90}
     \tikzstyle{gr}=[draw,circle,green!50!black,fill=green!50!black,scale=.6]
     \tikzstyle{Bl}=[draw,circle,blue,scale=.6]
     \tikzstyle{R}=[draw,circle,fill=red,scale=.6]
     \tikzstyle{bl}=[draw,circle,fill=black,scale=.35]
     \tikzstyle{bbc}=[draw,circle,fill=black,scale=.75]
     \tikzstyle{bbcs}=[draw,circle,fill=black,scale=.5]
     \tikzstyle{rc}=[circle,fill=red,scale=.6]
     \tikzstyle{wc}=[draw,circle,scale=.75]
\usepackage{array} 
\usepackage{multirow} 
\usepackage{colortbl} 
\usepackage{arydshln} 
\usepackage{stfloats}  
\usepackage{cellspace}
\usepackage{cellspace}
\usepackage{xcolor}   

\usepackage{url} 

\newcommand{\beq}{\begin{equation}}
\newcommand{\eeq}{\end{equation}}
\newcommand{\xdasharrow}[2][->]{
\tikz[baseline=-\the\dimexpr\fontdimen22\textfont2\relax]{
\node[anchor=south,font=\scriptsize, inner ysep=1.5pt,outer xsep=2.2pt](x){#2};
\draw[shorten <=3.4pt,shorten >=3.4pt,dashed,#1](x.south west)--(x.south east);
}
}
\def\del{{\partial}}
\def\bar{\overline}
\def\delb{{\bar\del}}
\def\til{\widetilde}
\def\hat{\widehat}
 
\def\vev#1{{\langle{#1}\rangle}} 
\newcommand{\bpmat}{\begin{pmatrix}}
\newcommand{\epmat}{\end{pmatrix}}
\newcommand{\bsmat}{\begin{smallmatrix}}
\newcommand{\esmat}{\end{smallmatrix}}

\def\^{\wedge}

\def\Tr{\mathop{\rm Tr}}
\def\Im{\mathop{\rm Im}}

\def\U{\mathop{\rm u}}
\def\SU{\mathop{\rm su}}

\def\SO{\mathop{\rm so}}
\def\SL{\mathop{\rm SL}}
\def\GL{\mathop{\rm GL}} 
\def\Sp{\mathop{\rm sp}}

\def\C{\mathbb{C}} 

\def\N{\mathbb{N}} 
\def\P{\mathbb{P}} 
\def\R{\mathbb{R}} 
\def\Z{\mathbb{Z}} 
\def\ff{{\mathfrak f}}

\def\bm{{\bf m}}

\def\ub{{\bar u}}

\def\cC{{\mathcal C}}
\def\cCb{\overline\cC}
\def\cCh{{\hat\cC}}

\def\cE{{\mathcal E}}

\def\cI{{\mathcal I}}
\def\cJ{{\mathcal J}}
\def\cK{{\mathcal K}}
\def\cL{{\mathcal L}}
\def\cM{{\mathcal M}}
\def\cN{{\mathcal N}}

\def\a{{\alpha}}

\def\ba{{\boldsymbol\a}}

\def\g{{\gamma}}

\def\d{{\delta}}

\def\D{{\Delta}}
\def\e{{\epsilon}}

\def\z{{\zeta}}

\def\k{{\kappa}}
\def\l{{\lambda}}
\def\L{{\Lambda}}
\def\m{{\mu}}

\def\n{{\nu}}
\def\x{{\xi}}
\def\r{{\rho}}
\def\s{{\sigma}}

\def\t{{\tau}}

\def\f{{\phi}}
\def\vf{{\varphi}}

\def\up{{\upsilon}}
\def\w{{\omega}}

\title{Seiberg-Witten geometries for Coulomb branch chiral rings which are not freely generated}
\author{Philip C. Argyres,}
\author{Yongchao L\"u}
\author{and Mario Martone}
\affiliation{University of Cincinnati,
Physics Department, PO Box 210011, Cincinnati OH 45221}
\emailAdd{philip.argyres@gmail.com}
\emailAdd{lychaoaa@gmail.com}
\emailAdd{martonmo@ucmail.uc.edu}

\abstract{Coulomb branch chiral rings of $\cN=2$ SCFTs are conjectured to be freely generated. While no counter-example is known, no direct evidence for the conjecture is known either. We initiate a systematic study of SCFTs with Coulomb branch chiral rings satisfying non-trivial relations, restricting our analysis to rank 1. The main result of our study is that (rank-1) SCFTs with non-freely generated CB chiral rings when deformed by relevant deformations, always flow to theories with non-freely generated CB rings. This implies that if they exist, they must thus form a distinct subset under RG flows.  We also find many interesting characteristic properties that these putative theories satisfy which may behelpful in proving or disproving their existence using other methods.}
 
\begin{document}
\maketitle

\section{Introduction and summary of results}

Four dimensional $\cN=2$ supersymmetric theories form a richly accessible corner of the space of quantum field theories which continues to yield new exact results even more than two decades after the pioneering work of Seiberg and Witten \cite{sw1,sw2}. For instance, the Coulomb branch (CB) of $\cN=2$ superconformal theories (SCFTs) has to be a scale invariant special K\"ahler (SK) variety, and the systematic study of these geometries and their behavior under mass deformations has allowed a complete classification of all $\cN=2$ SCFTs with a one-complex-dimensional CB \cite{paper1, paper2, allm1602, paper3, am1604}.  This classification not only assumed that the CB is one-dimensional but also planar --- that is, holomorphically equivalent to $\C$ as a complex manifold.  This assumption is made chiefly because all known CBs are equivalent to $\C^n$ as complex manifolds, where $n$ is the rank of the given theory.  In this paper we show that lifting the planar assumption produces many new, internally consistent, scale invariant SK geometries with consistent IR physical interpretations.  These geometries could be interpreted as CBs of conjecturally new SCFTs, although we are not able to either prove or disprove that such SCFTs exist. 

The CB of an $\cN=2$ SCFT can also be described algebraically.  In fact there is a particular class of superconformal multiplets, satisfying specific BPS shortening conditions, which generate the CB chiral ring \cite{Beem:2014zpa} and whose vevs parametrize the CB.  These multiplets will play a central role in our analysis and so it is worthwhile to define them right away.  They are called \emph{chiral} multiplets and denoted by $\cE_r$, where $r$ labels the $U(1)_r$ charge.  They are singlets under the remaining $SU(2)_R$ R-symmetry and their scaling dimensions are $\D(\cE_r)=r$.  Because the $U(1)_r$ charge is conserved and additive, the (lowest complex scalar components of the) $\cE_r$ form a chiral ring,
\begin{align}\label{CBchiralring}
\cE_r \cE_s = \sum_t c_{rst} \cE_t,
\end{align}
in the limit when they are taken to be coincident operator insertions.  

It is believed that the CB chiral ring of four dimensional $\cN=2$ SCFTs is freely generated, that is the generators of \eqref{CBchiralring} do not satisfy any non-trivial relations.  This is an algebraic way of saying that the space of their vevs is $\C^n$ as a complex manifold, where $n$ counts the number of independent generators, which corresponds to the rank of the theory mentioned above.  Despite the fact that, to our knowledge, there is no fundamental reason to believe that all CB chiral rings should be freely generated, the number of known examples and lack of counter-examples has motivated the conjecture that all $\cN=2$ SCFTs CB chiral rings are freely generated \cite{Tachikawa:2013kta, Beem:2014zpa}.  In this paper we start the systematic study of SCFTs with non-freely generated CB chiral rings in order to examine the content of this conjecture. 

Our analysis, which is restricted to the rank-1 case, is not able to either prove or disprove the conjecture.  Yet we provide a plausible explanation of why, if theories with non-freely generated CB chiral rings existed, we have not yet encountered any examples.  Also we find a series of remarkable results which highlight some interesting and novel properties of the CBs of theories with non-freely generated CB chiral rings.  We summarize these findings in the rest of this section while keeping the presentation as non-technical as possible; we leave the technical details for the remainder of the paper.

\subsubsection*{Non-planar topology}

\begin{figure}[ht]
\centering
\begin{tikzpicture}
\begin{scope}[scale=.9]
\node at (0,-3.5) {\bf (a)};
\draw[thick,draw=black!45] (0,0) -- (1.5,-2.5) arc (375:165:1.5cm and .5cm) -- cycle;
\draw[thick,dashed,black!25] (1.5,-2.5) arc (15:165:1.5cm and .5cm);
\draw[rotate=120,thick,draw=black!45] (0,0) -- (1.5,-2.5) arc (375:165:1.5cm and .5cm) -- cycle;
\draw[rotate=120,thick,dashed,black!25] (1.5,-2.5) arc (15:165:1.5cm and .5cm);
\draw[rotate=240,thick,draw=black!45] (0,0) -- (1.5,-2.5) arc (375:165:1.5cm and .5cm) -- cycle;
\draw[rotate=240,thick,black!45] (1.5,-2.5) arc (15:165:1.5cm and .5cm);
\node[R] (br) at (0,0) {};
\begin{scope}[xshift=5cm]
\node[single arrow, draw, black] at (0,0) {{\small deformation}};
\end{scope}
\begin{scope}[xshift=10cm]
\node at (0,-3.5) {\bf (b)};
\draw[thick,draw=black!45] (1.45,-0.03) .. controls (.6,-.5) .. (.75,-1.25) -- (1.5,-2.5) arc (375:165:1.5cm and .5cm) -- (-.75,-1.25);
\draw[thick,dashed,black!25] (1.5,-2.5) arc (15:165:1.5cm and .5cm);
\draw[rotate=120,thick,draw=black!45] (1.45,-0.03) .. controls (.75,-.6) .. (.75,-1.25)-- (1.5,-2.5) arc (375:165:1.5cm and .5cm) -- (-.75,-1.25);
\draw[rotate=120,thick,dashed,black!25] (1.5,-2.5) arc (15:165:1.5cm and .5cm);
\draw[rotate=240,thick,draw=black!45] (1.45,-0.03) .. controls (.5,-.4) .. (.75,-1.25)-- (1.5,-2.5) arc (375:165:1.5cm and .5cm) -- (-.75,-1.25);
\draw[rotate=240,thick,black!45] (1.5,-2.5) arc (15:165:1.5cm and .5cm);
\draw[rotate=20,thick,draw=black!45] (0.6,.2) arc (375:165:.5cm and .2cm);
\draw[rotate=20,thick,draw=black!45] (0.5,.1) arc (15:165:.4cm and .15cm);
\node[R] (br1) at (0,-1) {};
\node[R] (br2) at (1,.8) {};
\node[R] (br3) at (-1,.25) {};
\node[R] (br4) at (-.25,.6) {};
\end{scope}
\end{scope}
\end{tikzpicture}
\caption{(a) A scale-invariant CB which is a bouquet of 3 cones: the conformal vacuum is the common tip of the cones, while at all other points conformality is spontaneously broken.  (b) A ``near" deformation of the scale-invariant geometry, in which the conformal singularity splits into several conical or cusp-like metric singularities (the red dots) without affecting the asymptotically far geometry, while the complex geometry desingularizes into that of a 3-punctured genus-1 Riemann surface.}
\label{fig-cone}
\end{figure}
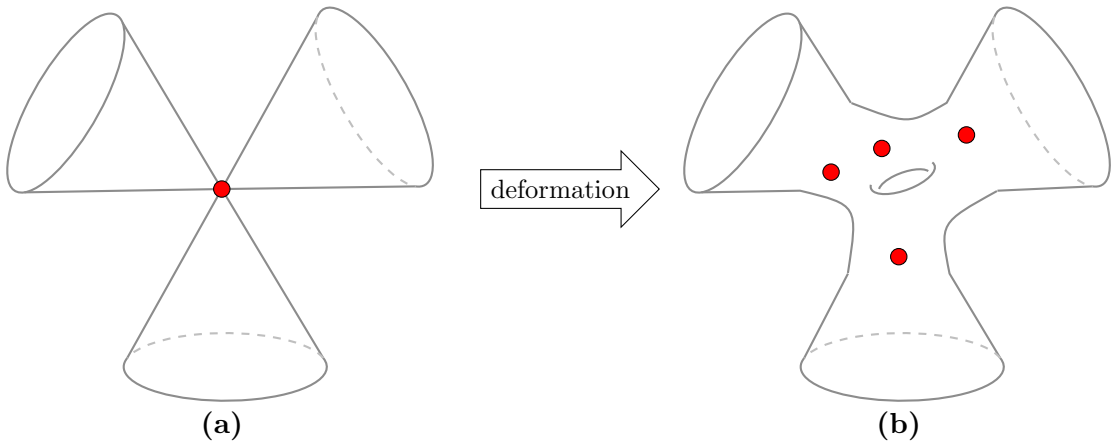

The first result that is worth highlighting is the connection between non-freely-generated CB chiral rings and non-planar CBs.
The CB of an $\cN=2$ SCFT with a non-freely-generated CB chiral ring under mass deformation becomes a non-compact genus $g$ Riemann surface with metric singularities which have the standard interpretation as loci on the CB where extra massless states enter the theory.  
This kind of CB geometry is depicted in figure \ref{fig-cone}b where the red dots represent metric singularities.

In the conformal limit the CB must be a SK scale invariant variety. In section \ref{2.1}, and in more detail in appendix \ref{irregular}, we show that the only SK scale invariant geometries have components which are cones with particular metric singularities at their tips.  Thus the most generic scale invariant CB geometry is a bouquet\footnote{A.k.a.\ wedge sum.} of $N$ cones sharing a common tip; see figure \ref{fig-cone}a. The case $N>1$ can only be realized if the CB chiral ring is non-freely generated although the converse is not true.  In appendix \ref{appA}, which collects explicit examples of non-freely-generated CB chiral rings, we work out cases with both $N=1$ and $N>1$. 

Let us now work out an explicit, yet very simple, example.  Consider a theory with a CB chiral ring generated by two operators $\cE_{6/5}$ and $\cE_{9/5}$ satisfying a single relation:
\beq\label{CBex0}
\cE_{6/5}^3=\cE_{9/5}^2 .
\eeq
Assume that both $\cE_{9/5}$ and $\cE_{6/5}$ can acquire a vev 
and call $u_1=\vev{\cE_{9/5}}$ and $u_2=\vev{\cE_{6/5}}$. Then the CB is described by the one complex dimensional variety in $\C^2$ satisfying
\beq\label{CBex1}
\ff(u_1,u_2) := u_1^2-u_2^3 = 0 .
\eeq
Scale invariance implies that the metric on this space is the metric of a flat cone.  This is a case where the CB spanned by a non-freely-generated chiral ring describes a single cone.  A first difference from the freely-generated case is that the plane curve $\ff(u_1,u_2)=0$ is singular, as a complex manifold, at $u_1=u_2=0$.

Turning on all relevant $\cN=2$ supersymmetry-preserving deformations does not lift the CB but only deforms it; see, e.g.,  \cite{paper1} for an in-depth discussion on this point.  In particular the CB of the deformed theory is still a plane curve which can be obtained by adding terms to \eqref{CBex1} which depend polynomially on both $u_1$ and $u_2$ and the relevant couplings. The possible relevant couplings are either mass terms or chiral deformation parameters \cite{paper1}.  Using some elementary deformation theory,\footnote{See, for example, \cite{Zoladek, ebeling, GLS07, AG-ZV1, AG-ZV2} for expositions of deformation theory, and also \cite{Xie:2015rpa, Chen:2016bzh, Wang:2016yha} for a discussion in a related context.} we can write down the generic deformation of \eqref{CBex1} as
\beq\label{CBex1D}
u_1^2 = u_2^3 + \l_2 u_2 + \l_1
\eeq
where $\l_1$ and $\l_2$ are holomorphic functions of the relevant couplings in the theory.

For generic values of $\l_1$ and $\l_2$ this deformed CB is nowhere singular as a complex manifold.  We readily compute the geometric genus of the one-dimensional complex space described by \eqref{CBex1D} using, e.g., the genus-degree formula, $g=\frac12(d-1)(d-2)$, where $d$ is the degree of the non-singular plane curve.  (Alternatively, we simply recognize \eqref{CBex1D} as the equation of an elliptic curve in Weierstrass form.)  Thus \eqref{CBex1D} describes a non-planar CB which is a non-compact genus 1 Riemann surface --- i.e., it is missing the point at $(u_1,u_2)=(\infty,\infty)$.

This example generalizes straightforwardly.  Appendix \ref{appB} collects several examples of deformed CBs which have the complex structure of an $N$-punctured genus-$g$ Riemann surface, $\cM_{g,N}$.  In the limit where the deformation parameters are taken to zero, $\cM_{g,N}$ degenerates to the bouquet of $N$ once-punctured spheres described above.  Note that although $\cM_{g,N}$ is a complex (smooth) manifold, it has \emph{metric} singularities corresponding to vacua where additional states charged with respect to the low energy $\U(1)$ gauge theory become massless; these are the red dots in figure \ref{fig-cone}b.  

\subsubsection*{Apparent violation of the unitarity bound}

Unitarity of four dimensional conformal field theories restricts the scaling dimension of scalar operators, $\f_i$, to $\D(\f_i)\geq1$. When the CB coordinates, $u_i$, are the vevs of operators in the CFT spectrum, $u_i =\vev{\f_i}$, this constraint implies $\Delta(u_i)\geq1$.  But in the presence of non-trivial relations among the CB chiral ring generators, the appropriate coordinates to parametrize the CB are generically \emph{not} the vevs of any operator in the spectrum and thus their scaling dimension can be smaller than one.

Consider the SCFT CB example \eqref{CBex0} analyzed above.  The single relation in $\C^2$ carves out a one dimensional complex cone which, ignoring the singularity at the tip, can be parametrized by a single complex coordinate $u\in\C^*$.  We call $u$ the \emph{uniformizing parameter} of the cone (minus its tip).  What is the relation between $u$ and $(u_1,u_2)$, the vevs of the two CB generators?  By scale invariance $u=u_1^{\g}=u_2^{3\g/2}$, where $\g$ is fixed by the requirement that $u$ is a good (i.e., single-valued) coordinate near the origin.  In particular, upon looping around the origin, $u\to e^{2i\pi}u$, $(u_1,u_2)$ transform to $(e^{i2\pi/\g}u_1, e^{i4\pi/3\g}u_2)$.  Their single-valuedness implies that $\g=1/3$, so that $u=u_1^{1/3}=u_2^{1/2}$.  

Since the scaling dimension of (the scalar primary of) an $\cE_r$ multiplet is $r$, the scaling dimension of the uniformizing parameter of the CB associated to the chiral ring in \eqref{CBex0} is
\beq\label{CBrel2}
\D(u) = \frac13 \D(u_1) = \frac12 \D(u_2) = \frac35 
\qquad\Rightarrow\qquad \D(u)<1.
\eeq
In this way a CB associated with a non-freely generated CB chiral ring can apparently violate the unitarity bound, without actually doing so. 

\subsubsection*{Irregular singularities}

This observation has profound implications for the set of CB geometries allowed when there are non-freely-generated CB chiral rings.  As discussed in detail in section \ref{2.1} and appendix \ref{irregular}, once we drop the restriction that $\D(u)\geq1$, there emerges an infinite set of allowed scale invariant SK geometries, even when restricting to the rank-1, or one complex dimensional, case.  In fact the well-known seven scale invariant conical geometries --- usually denoted $II$, $III$, $IV$, $I_0^*$, $IV^*$, $III^*$, $II^*$ following the Kodaira classification of singular elliptic fibers \cite{KodairaI,KodairaII} ---
are now just the $m=0$ members of an infinite series of conical geometries parametrized by non-negative integers $m$.  We will refer to them by their Kodaira labels together with an $(m)$ superscript; their properties are recorded in table \ref{table:eps}. 

For $m>0$ the scaling dimension of the uniformizing parameter of these geometries is less than 1 and thus these CBs, in light of the discussion above, can only be associated to SCFTs with non-trivial CB chiral rings.  We call these $m>0$ geometries \emph{irregular type singularities}.

More interestingly, the non-scale-invariant $I_n^{(m)}$ and $I_n^{*(m)}$ geometries all have a low energy $\U(1)$ gauge coupling which becomes arbitrarily weakly coupled as one approaches the singularity.  The $m=0$ versions of these geometries are just the CBs of familiar $\U(1)$ and $\SU(2)$ IR-free gauge theories.  The $m>0$ geometries would therefore seem to describe IR-free theories with non trivial CB relations.  There is no known example of such a theory, and one might expect that their existence could be excluded because their CBs are weakly coupled.  But, perhaps surprisingly, such low energy effective actions with non-trivial CB relations can arise as quantum-corrected CB chiral rings in a fairly natural way from RG flows from UV SCFTs with non-trivial CB chiral rings.  This is discussed in section \ref{s:dis}.  Thus we find no fundamental obstacle to the possible existence of such theories and their associated CB geometries.

\subsubsection*{Main result of the paper}

Our main result will be to show that SCFTs with non-freely-generated rank-1 CBs necessarily flow to exotic fixed point theories with non-freely-generated CB chiral rings and therefore form a subset of $\cN=2$ field theories which are not connected by RG flows to the rank-1 theories with regular, planar CB geometries, classified in \cite{paper1, paper2, allm1602, paper3, am1604}. This will be done showing that for rank-1 theories if, and only if, there are non-trivial SCFT CB chiral ring relations, then at least one of the metric singularities of the geometry of its deformed CB must be of irregular type, i.e., with $m>0$ in table \ref{table:eps}. These singularities, as we described above, can only be interpreted as fixed points with non-freely generated CB chiral rings.

We argue this in section \ref{s:ru}, and subsection \ref{ssGA} in particular, by first showing that if a deformation makes the CB a smooth complex manifold, as in \eqref{CBex1D}, the CB is holomorphically equivalent as complex manifold to a compact Riemann surface, $\cCh_{g,N}$, of genus $g$ with $N$ marked points.  This follows from an examination of the possible SK geometries at the $N$ asymptotic infinities.  It then follows from an application of the Riemann-Roch theorem to $\cCh_{g,N}$, together with an analysis of how the discriminant of the SW elliptic curve encoding the SK geometry of the CB behaves near irregular singularities, that there must be an irregular singularity whenever $g>0$ or $N>1$.  We then show that even if the deformed CB is not smooth as a complex manifold, there must be at least one IR fixed point on it which has a non-freely-generated CB chiral ring if the UV fixed point theory also has a non-trivial CB chiral ring.

\subsubsection*{Existence of non-planar SK geometries}

Finally, all the above considerations do not actually construct an example of a family of deformed CBs with non-planar topology and  SK geometries.   In particular, we only demanded a SK structure locally, and then found constraints from assuming a global complex structure, but did not try to apply the more stringent condition that a SK structure should also exist globally on the CB.  This leaves open the possibility that there may, in fact, be no non-planar rank-1 SK CB geometries.  

In section \ref{ssCC} we close this door by showing how to construct a large class of consistent SK CB geometries as non-planar Riemann surfaces which are multi-sheeted covers of planar SK CB solutions and there are even more examples that can be constructed by modifications of this procedure. We conclude that CBs associated to SCFTs with non-freely generated CB chiral rings are not physically inconsistent yet are peculiar in many ways.

\bigskip

The outline of the rest of the paper is as follows.  

Section \ref{s:sir1} discusses the connection of the CB chiral ring of SCFTs with the geometry of their scale invariant rank-1 CBs.   Appendix \ref{appA} collects some examples of the kinds of algebraic behaviors which are possible with rank-1 chiral rings, and appendix \ref{irregular} shows how to construct SK geometries on these CBs.  We also briefly discuss in section \ref{s2.3} the confrontation of these examples with recent results from the $\cN=2$ bootstrap.

Section \ref{s:ru} turns to the discussion of the non-scale-invariant  rank-1 CB geometries that can result from the deformation of the scale-invariant geometries described in section \ref{s:sir1}.  In particular, we work out the form of the SW curves in the vicinity of the various possible SK singularities, and then patch them together globally on the CB to find the results described above.  Appendix \ref{appD} shows how to construct algebraic SW curves for the scale-invariant CBs of section \ref{s:sir1}, while appendix \ref{appB} provides some examples of miniversal deformations of those CBs.

We conclude in section \ref{s:dis} with a discussion of the physical interpretation of irregular singularities with IR-free CB couplings.  We also mention a few outstanding open questions.

\section{Rank-1 CBs of SCFTs}
\label{s:sir1}

In lagrangian $\cN=2$ conformal theories, the CB is the component of the moduli space spanned by vevs of operators given by gauge invariant combinations of the scalar component of the vector multiplet. These operators are superconformal primaries of the multiplets introduced above satisfying specific shortening conditions and, following the terminology introduced in \cite{Dolan:2002zh}, are indicated as $\cE_r$. The straightforward non-lagrangian generalization is to think of the CB as the space of vevs of the chiral ``$\cE_r$" scalar primaries of the SCFT.\footnote{See, e.g., appendix B of \cite{paper1} for a summary of the unitary positive-energy representations of the $\cN=2$ superconformal algebra.}  The $\cE_r$ fields form a (commutative) chiral ring,\footnote{We will use a ``0" superscript to mark objects associated to a SCFT, and will drop it once we break conformal invariance explicitly by turning on deformations.} 
\begin{align}\label{CBchr}
R^0_\text{CB} = \C\{\f_1,\f_2,\ldots\}/\cI^0 ,
\end{align}
where the $\f_i$ are some set of generating fields, $\C\{\f_1,\f_2,\ldots\}$ is the ring of analytic function in $\f_1,\f_2,\ldots$ and $\cI^0$ is the ideal of relations among those generators, itself generated by some set of relations
\begin{align}\label{CBpolys}
P^0_k(\f_i) = 0 .
\end{align}
The chiral ring is graded by the scaling dimensions, $\D(\f_i) \in \R$, (proportional to the $\U(1)_R$-charges) of the $\f_i$.  So the generating functions $P^0_k$ of $\cI^0$ are each homogeneous in dimension.

Scalar fields of unitary 4d CFTs have dimension greater than or equal to 1 \cite{Mack:1975je}.  Homogeneity and analyticity of the chiral ring relations then imply that the  $P^0_k$ are all polynomial in the $\f_i$.  Thus, without loss of generality we can take $R^0_\text{CB} = \C[\f_1,\f_2,\ldots]/\cI^0$ in place of \eqref{CBchr}, i.e., a polynomial ring, not just an analytic one.

Under the
\begin{align}\label{CBass}
\text{{\bf Assumption:} that all $\cE_r$ fields in an $\cN=2$ SCFT are potential flat directions,} 
\end{align}
i.e., that there exist vacua with spontaneously broken conformal invariance for arbitrary non-zero vevs of the $\f_i$ fields compatible with their chiral ring relations, then the CB is the algebraic set,
\begin{equation}\label{CB}
\cCb^0 =  V(\cI^0) , 
\end{equation}
defined as the common zeros of the elements of $\cI^0$ thought of as complex functions of the $\f_i$. If this assumption is incorrect, then some of the $\cE_r$ fields do not get vevs, and the CB is presumably $V(\cJ)$ for some ideal $\cJ \supset \cI^0$.

If $\cI^0=\{\varnothing\}$ or there is a basis of $R^0_\text{CB}$ for which all the relations can be solved for and thus $R^0_\text{CB}$ can be written in terms of a smaller set of generators satisfying no non-trivial relations among them, the chiral ring is freely generated. It has been conjectured \cite{Tachikawa:2013kta,Beem:2014zpa} that the CB ring of $\cN=2$ SCFTs is always freely generated. Solid evidence for this conjecture is provided by the study in \cite{Chacaltana:2016shw,Chacaltana:2017}, providing a zoo of CFTs with seemingly non-freely generated chiral ring CBs, but which can always be shown to be freely generated. The purpose of this paper is to explore CB geometries whose coordinate rings are not freely generated and understand from first principles whether they are allowed.

In particular we start by exploring the relation between $R^0_\text{CB}$, the CB chiral ring, and the CB geometry. We  restrict our study to one-dimensional CB chiral rings, that is, we describe the most general CB chiral rings and geometries of rank-1 $\cN=2$ SCFTs.  The regular points of the CB form a complex, special K\"ahler (SK), manfold. We first show that each connected component is holomorphically equivalent to $\C^*$ as a complex manifold and has the metric of a flat cone (minus the tip).  We then argue that superconformal invariance implies that the CB is a bouquet of such cones, i.e., the disjoint union of a set of these cones modulo the identification of their tips.  We then discuss the relationship between the chiral ring of the SCFT and the geometry of the CB, and how this is reflected in the (singular) structure of the CB.

\subsection{Global structure of the CB}\label{sec2.4}

We now describe the geometric structure of the CB which we have thus far described only as an algebraic variety.  Denote by $\cCb^0$ the Coulomb branch.  It is a complex variety with a special K\"ahler metric.  It can have both singularities as a complex manifold, as well as places where the metric becomes singular.  These two different kinds of singular loci need not be the same.  The  locus where $\cCb^0$ fails to be a complex manifold has complex co-dimension one (or greater), so in the rank-1 case consists of a discrete set of points.   We will call this locus the complex singularities of $\cCb^0$, and the locus where the metric degenerates as the metric singularities.  Since the metric is K\"ahler, the metric singularities are also in complex codimension 1, so are also a discrete set of points in the rank-1 case.

We will now argue that for a SCFT the complex singularities and the metric singularities both consist of at most one point and if both exist they must coincide.   

Let $\cC^0$ be $\cCb^0$ minus both its complex and metric singularities.   It is thus a (smooth) complex manifold, though it may have multiple disconnected components.  There is no a priori reason that the various components of $\cC^0$ have to be equi-dimensional.  But in this paper we make the following simplifying
\begin{align}\label{CBrk1}
\text{{\bf Assumption:} all connected components of $\cC^0$ are 1-dimensional.}
\end{align}
In algebraic terms this implies that each irreducible component of $V(\cI^0)$ is one-dimensional.\footnote{Note that this assumption does not imply that all components of a deformed (non-conformal) CB need be 1-dimensional.  An example of a flat deformation of a 1-dimensional scale-invariant CB which has disconnected 0-dimensional components is given in appendix \ref{appB}.}

The 4d $\cN=2$ superconformal algebra contains an $\SO(1,1)_D \times \U(1)_R$ dilatation and R-symmetry.   These combine to give a non-trivial holomorphic $\C^*$ action on the chiral ring of $\cE_r$ scalar primaries of the SCFT, and therefore such an action also extends to $\cC^0$.  A connected 1-complex-dimensional manifold with a non-trivial holomorphic $\C^*$ action without a fixed point is holomorphically equivalent to $\C^*$ itself, i.e., a punctured complex plane \cite{Akhiezer:1995}.  Say there are $N$ connected components, $\cC^0_a$, $a=1,\ldots,N$, so that $\cC^0$ is their disjoint union.   We will coordinatize $\cC^0_a\simeq \C^*$ as the complex $u_a$-plane minus the point $u_a=0$, and we take the $\C^*$ action to be simply $u_a \mapsto \l u_a$ for $\l\in\C^*$.   This complex $u_a$ coordinate is the {\it uniformizing parameter} for $\cC^0_a$ introduced in the introduction.  

Furthermore, since the $\C^*$ action is a complex homothety of the metric on $\cC^0$,  it follows (see appendix \ref{irregular}) that each connected component has the geometry of a flat cone minus the tip.  In particular, none of the points of $\cC^0$ are metric singularities and the tips, $u_a=0$, of each $\cC^0_a$ component cone is at finite distance.  Denote by $\cCb^0_a$ the closure of $\cC^0_a$ to include the point at $u_a=0$.

The tip of the $\cCb^0_a$ cone is a conformal vacuum of the SCFT, since it is a fixed point of dilatations.  Since it is at finite distance from points in $\cC^0$ in the low energy effective action metric, it should be included in the CB of vacua.  

Assuming that the CB $\cCb^0$ is connected, then by scale invariance there can only be one fixed point of the $\C^*$ action, since if there were more then the metric distance between them would provide a scale.  Thus the assumption that the CB is connected is the same as the assumption that the SCFT has a unique conformal vacuum. 

With this assumption, we have found that the general rank-1 CB, $\cCb^0$, of a SCFT is the disjoint union of a collection of such cones with their tips identified, as pictured in figure \ref{fig-cone}a.  Thus it is the bouquet of cones 
\begin{align}\label{}
\cCb^0 = \bigvee_{a=1}^N \cCb^0_a .
\end{align}
Note that each conical branch, $\cCb_a$, will have its own uniformizing parameter $u_a$ with a priori different conformal dimensions $\D(u_a)$.  The only point of $\cCb^0$ which can be either a metric or complex singularity is the common tip of the cones.  When $N>1$ then it is both, but when $N=1$ it is possible for the tip to be a metric singularity but not a complex singularity, as we will see in the next subsection.

\subsection{Local CB geometry}\label{2.1}

We now characterise the possible ways the metric can degenerate at the tip of the bouquet of cones that make up the CB.   In fact, we list the countably infinite number of distinct metrics that are allowed by demanding a special K\"ahler (SK) structure on the CB.   In contrast, we are not able to characterize the uncountably infinite number of distinct complex singularities that can occur at the tip.

We study the metric singularities at the tip by looking at the way the metric can degenerate on a single conical component, $\cCb^0_a$, of the CB.  The general metric degeneration will then just be a separate choice of these possible metric degenerations for each component.

\paragraph{Irregular geometries.} 

The special K\"ahler geometry of the $\cCb^0_a$ cone can be described \cite{sw2} by an elliptic fibration, $X\to \cCb^0_a$, whose fibers, $X_u$ for $u\in\cCb^0_a$, are thus genus-1 Riemann surfaces which vary holomorphically over $\cCb^0_a$.  We show in appendix \ref{irregular} that the SK condition at the regular points of $\cCb^0_a$ imply that only a discrete, but infinite, set of geometries is allowed.  These are listed in table \ref{table:eps}.
(In appendix \ref{irregular} and in table \ref{table:eps} we have also included two series, denoted $I^{*(m)}_n$ and $I^{(m)}_n$, of non-scale-invariant cusp-like singularities which are also allowed by the SK conditions.  Though these cannot occur as the CB of a SCFT (since they are not scale invariant), they will play an important role as possible IR singularities once we break scale invariance by turning on relevant deformations in section \ref{s:ru}.)

\begin{table}
\centering
$\begin{array}{|c|c|c|c|c|}
\hline
\multicolumn{5}{|c|}{\text{\bf Possible rank 1 SK cones and cusps}}\\
\hline\hline
\ \text{name}\ \,&\ \text{opening angle}\ \,&\ \D(u)\ \, & M_0               & \t_0  \\
\hline
II^{*(m)} 
& 2\pi (m+\frac16) & \frac6{1+6m} & ST & e^{i\pi/3}  \\
III^{*(m)} 
& 2\pi (m+\frac14) & \frac4{1+4m} & S   & i                    \\
IV^{*(m)} 
& 2\pi (m+\frac13) & \frac3{1+3m} & -(ST)^{-1} & e^{2i\pi/3}\\
I_0^{*(m)} 
& 2\pi (m+\frac12) & \frac2{1+2m} & -I   &\ \text{any}\ \t\ \, \\
IV^{(m)} 
& 2\pi (m+\frac23) & \frac3{2+3m} & -ST & e^{2i\pi/3}     \\
III^{(m)} 
& 2\pi (m+\frac34) & \frac4{3+4m} & S^{-1}  & i               \\
II^{(m)} 
& 2\pi (m+\frac56)& \frac6{5+6m} & (ST)^{-1} & e^{i\pi/3} \\
I_0^{(m)} 
& 2\pi (m+1)        & \frac1{1+m}  & I    & \text{any}\ \t \\[1mm]
\hdashline
I_n^{*(m)} 
& 0\ (\text{cusp})\ & \frac2{1+2m} & -T^n  & i\infty         \\
I_n^{(m)} 
& 0\ (\text{cusp})\ & \frac1{1+m}  & T^n    & i\infty   \\[1mm]
\hline
\end{array}$
\caption{All geometries of 1-dimensional special K\"ahler cones and cusps.  Shown are the allowed values of the opening angle $2\pi\e$, of the mass dimension of the uniformizing parameter $\D(u)$, of the EM duality monodromy conjugacy class about the tip of the cone $M_0$, and of the value of the complex modulus of the fiber near the tip $\t_0$.  $m\ge0$ is a non-negative integer, and $n>0$ is a positive integer.  The first eight rows are scale-invaraint conical geometries, while the last two rows are non-scale-invariant cusp-like geometries.\label{table:eps}}
\end{table}

The study\footnote{See, e.g., \cite{paper1} for a detailed discussion.} of scale invariant CB geometries whose uniformizing parameter, $u$, satisfies the (naive) unitarity condition that $\D(u)\ge1$, found only a finite number of such geometries, in correspondence with a subset of Kodaira's classification \cite{KodairaI,KodairaII} of possible singular elliptic fibers of elliptic surfaces.  These correspond to the eight scale-invariant ``$m=0$" conical geometries of table \ref{table:eps}.  In appendix \ref{irregular}, by applying the SK condition without imposing the naive unitarity condition, we find instead that the list of possible scale-invariant rank-1 geometries is far larger.  In fact, for each Kodaira type we now have an infinite series of possible geometries labelled by a non-negative integer $m$.   The infinite series of $m\neq0$ geometries have uniformizing parameters with $0<\D(u)<1$.  We will call these $m\neq0$ geometries \emph{irregular geometries}.

To our knowledge this is the first time that the existence of these irregular geometries has been pointed out in the physics literature.  This is not surprising.  In fact, as explained in the introduction, irregular geometries are unphysical if we assume that the CB chiral ring is freely generated since in such cases $u$ is identified as the vev of the scalar primary of a chiral operator $\cE_r$ in the SCFT operator algebra.  But scale invariance and unitarity in 4 dimensions imply \cite{Mack:1975je} that the scaling dimension of a scalar field is greater than or equal to 1.  Thus from unitarity,
\begin{align}\label{unitarity}
u = \vev\f \qquad \Rightarrow \qquad
\D(u)\ge 1 \quad \text{and so} \quad m=0 
\quad \text{in table \ref{table:eps}.}
\end{align}
The situation changes if there are relations in the CB chiral ring, since the uniformizing parameter is in general no longer the vev of any scalar operator in the SCFT and thus \eqref{unitarity} no longer applies.  A simple example, see \eqref{CBrel2}, illustrating this has been already given in the introduction; many more are given in appendix \ref{appA}.

\paragraph{Complex singularities.} 

The tip of the bouquet of cones making up $\cCb^0$ is a complex singularity if and only if the CB chiral ring is not freely generated.  This can be seen explicitly from the description of the CB as an algebraic variety in \eqref{CB}.  The relations in $\cI_0$ are polynomials, $P_k^0(\f_i)$, weighted homogeneous in terms of the dimensions of the $\f_i$.  If any $P_k^0$ has $\f_i$ alone as one of its terms, then $\f_i$ can be eliminated from the description of the chiral ring in favor of the others.  After eliminating all such $\phi_i$, none of the remaining $\phi_i$ appears alone in any term of the remaining $P_k$, or all the $P_k$ are identically satisfied.  In the latter case the CB is freely generated.  The former case implies that $dP_k^0(\f_i)|_{\f_i=0}=0$, and thus the algebraic variety described by $P_k^0(\f_i)=0$ is singular at $\{\f_i=0\}$, the fixed point of the $\C^*$ complex scaling action, i.e., at the tip of the bouquet of cones.  

If the CB chiral ring is freely generated, then it has no complex singularity, and its uniformizing parameter is just the vev of the generating field, $u = \vev\f$.  Thus in this case, by \eqref{unitarity}, the geometry of the CB cone must be one of the eight regular $m=0$ geometries listed in table \ref{table:eps}.  Each of these has a metric singularity at the tip of the cone, except for the $I^{(0)}_0$ geometry which is just that of the euclidean plane.

The converse of this statement does not hold, however.  Even if the CB geometry is that of one of the regular $m=0$ cones in table \ref{table:eps}, it does not follow that the CB chiral ring is freely generated.  Indeed, it is easy to generate examples;  for instance, the simple example given in \eqref{CBrel2} in the introduction describes an $m=0$ $II^{(0)}$ geometry if the two operators are taken to have dimensions $12/5$ and $18/5$ instead of $6/5$ and $9/5$.

In general, there is an enormous number of inequivalent chiral rings that describe the same metric geometry, but which differ only by the complex singularity at the tip of the cones.  To help the reader gain some experience with the possible relationship between the CB chiral ring and the CB geometry, we give some increasingly baroque examples of rank-1 homogeneous algebraic varieties in appendix \ref{appA}.

\subsection{Comparison to constraints from the superconformal bootstrap}\label{s2.3}

Even though $\cN=2$ SCFTs with non-freely generated CB chiral rings are largely unexplored, there are a few constraints on possible relations among CB chiral ring operators, or lack thereof, that come from the conformal bootstrap analysis.  The purpose of this section is to summarize them.

The $\cN=2$ super-conformal bootstrap program was outlined and initiated in \cite{Beem:2014zpa}.  When restricted to a plane, the generators of $\cN=2$ Higgs branch chiral rings are captured by a 2d-chiral algebra \cite{b+13,Beem:2014rza}.  This observation tremendously simplifies ``boostrapping'' the Higgs branch compared to a similar analysis of CB operators.  Nevertheless an analysis of $\vev{\cE_r \cE_r \bar\cE_r \bar\cE_r}$ four-point function has been carried out in \cite{Beem:2014zpa} and the more elaborate analysis including mixed correlators $\vev{\cE_{r_1} \cE_{r_2} \bar\cE_{r_1} \bar\cE_{r_2}}$ for general $r_1$ and $r_2$ in a following paper \cite{Lemos:2015awa}.

Using the numerical bootstrap it is possible to exclude nilpotent relations of the kind
\beq\label{CBrel}
\cE_{r_1}\cE_{r_2}=0
\eeq 
for general $r_1$ and $r_2$.  This is because the OPE of $\cE_{r_1}\cE_{r_2}$ contains the operator $\cE_{r_1+r_2}$ and \eqref{CBrel} implies the vanishing of the corresponding OPE coefficient $\lambda_{\cE_{r_1}\cE_{r_2}\cE_{r_1+r_2}}$.  Using numerical bootstrap techniques it is possible to bound from below $\lambda^2_{\cE_{r_1}\cE_{r_2}\cE_{r_1+r_2}}$, constraining it to be strictly non-zero for a range of values of $(r_1,r_2)$.  This can be turned in a bound on the absence of nilpotent relations.  The strongest result was obtained in \cite{Beem:2014zpa}
\beq
\cE_r^2\neq 0 \quad{\rm for} \quad r\leq2.6
\eeq
A slighter weaker bound for general $r_1$ and $r_2$ can be obtained by studying mixed correlators.  The results are reported in figure 10 of \cite{Lemos:2015awa}. 

Relations of the type $\cE_{r1}^{p_1} = \cE_{r2}^{p_2}$ could be plausibly probed by considering a mixed system of correlators involving also higher powers of $\cE_{r1}$ and $\cE_{r2}$. However, at the moment this analysis appears to be computationally very demanding.\footnote{We thank Madalena Lemos for discussions on this point.}

\section{Constraints from deformations of rank-1 CBs}
\label{s:ru}


It should be clear from our discussion thus far, if it wasn't a priori clear, that the possible richness of the non-freely-generated CB chiral ring structure is hidden at the conformal vacuum living at the tip of the bouquet of cones.  Since we do not have direct effective field theory methods to probe interacting conformal vacua, we will instead deform the CFT by local operators.  As long as those deforming operators preserve $\cN=2$ supersymmetry along the RG flows that they generate, they will deform but not lift the CB.   Our main questions then become:
\begin{enumerate}
\item How are aspects of the CB chiral ring (e.g., its minimal number of generators and relations, whether it is irreducible, its nilradical) reflected in the deformed special K\"ahler geometry?
\item Are there obstructions to the existence of special K\"ahler deformations which suggest restrictions on which kinds of CB chiral rings might be consistent?
\end{enumerate}
Unfortunately, we will not be able to answer either of these questions in generality.  But we will show how to construct many examples of consistent deformations of scale-invariant CBs with non-freely-generated CB chiral rings in the remainder of the paper.  These will at least give hints as to how chiral ring relations appear in the low energy moduli space geometry.

$\cN=2$-preserving deformations preserve a special K\"ahler (SK) structure at the regular points of the CB, that is they don't lift the CB but they only deform it \cite{paper1}. The study of CB deformations has been used extensively to learn about and constrain planar SCFTs \cite{sw1,sw2,paper1,paper2,allm1602,am1604,paper3}. In this section we will show that through a careful analysis of CB deformations in the non-planar case, we can also learn about possible SCFTs with non-freely generated CBs. In order to study the deformations of rank-1 CBs we will make use of the description \cite{sw1, sw2, Donagi:1995cf} of the total space, $X$, of the CB (a.k.a., the Seiberg Witten curve) as an elliptic fibration with a certain holmorphic 2-form.   

We start by determining the possible algebraic forms of the metric singularities allowed by the condition of SK geometry as well as the allowed asymptotic forms of the geometry at the $N$ metric infinities.  These agree (as they must) with the list of possible rank 1 SK cones and cusps found in the last section and listed in table \ref{table:eps}.  In particular, as emphasized above, there are in addition to the regular ($m=0$) singularities corresponding to the Kodaira classification of singular elliptic fibers, an infinite set of irregular ($m>0$) versions of each of these.

Then, as an intermediate step, useful for making general statements about $\cN=2$-preserving deformations, we assume that the generic deformation of the CB is a smooth complex manifold.  This implies that $\cCb$, the CB associated to the deformed SCFT, will be a regular Riemann surface, $\cCh_{g,N}$, with no manifold singularities (see section \ref{2.1}) and whose topology is characterized by its genus, $g$, and the number of punctures, $N$.
Using this and analyticity to patch together the possible local and asymptotic behaviors of rank-1 SK geometries will allow us to argue our main result, which is that rank 1 unitary SCFTs with non-freely-generated CB chiral rings flow to at least one IR fixed point with a non-freely-generated CB chiral ring.

But this is not the end of the story, for the CB deformation and the deformation of the elliptic fibration are subject to two further sets of consistency conditions that will be discussed in section \ref{ssCC}. In particular, it is not clear a priori that there are \emph{any} deformed CB geometries with irregular singularities and which obey these extra conditions. In subsection \ref{ssCC} we show how to construct many examples of deformations with irregular IR singularities which do satisfy them.   The possible field theory interpretation of such CB geometries will be discussed in section \ref{s:dis}.

\subsection{Seiberg Witten curve and form}
\label{ssSW}

Recall \cite{sw1,sw2} that rank-1 SK geometries can be described in a neighborhood of any regular point of the CB by a family of elliptic curves together with a choice of a meromorphic 1-form on the curve, varying holomorphically on the CB.  We assume that the  singularities of the CB are isolated and polar in nature which, we will see, seems to be necessary to ensure the existence of well-defined IR scaling behavior for all CB vacua.  Other than setting notation, the main purpose of this subsection is to point out the existence of an ambiguity in defining the Seiberg-Witten data describing rank-1 SK geometries which we call the \emph{analytic gauge freedom}, see \eqref{SK6} and \eqref{SK7}.  (To be clear, this freedom is not related to space-time gauge transformations.)  This ambiguity plays an important role, as we will discuss in the next section, in  understanding the global form of the CB geometry.

We denote the general deformed (non-scale-invariant) CB by $\cCb$, and the complex manifold of its (both complex and metrically) non-singular points by $\cC$.  Let $u$ be a local complex coordinate on $\cC$.  The total space of the SW curve fibered over $\cCb$ is denoted $X$.  We describe the genus-1 fibers, $X_u$, as elliptic curves in Weierstrass form,
\begin{align}\label{SK1}
X_u: \qquad y^2 = x^3 + f(u)\, x\, z^4  + g(u)\, z^6,
\end{align}
where $(z,x,y)$ are homogeneous coordinates in weighted $\P^2_{(1,2,3)}$ projective space, and $f$ and $g$ have complex analytic dependence on a local CB coordinate, $u$.  The modulus, $\t(u)$, of the complex structure of this elliptic curve is the low energy $\U(1)$ gauge coupling on the CB.  The curve $X_u$ has a singularity in its complex structure (at fixed $u$) when the discriminant,
\begin{align}\label{defdisc}
\text{disc} := - 4 f^3 - 27 g^2,
\end{align}
vanishes.  These singularities in the complex structure of $X_u$ appear as metric singularities on the CB.  We will assume that $f$ and $g$ are meromorphic (so have isolated singularities in $u$ which are just poles), and that disc does not vanish identically.  It then follows that the zeros of disc are also isolated.  We will argue below that with this assumption, all singularities of $f$ and $g$, including those at metric infinity, will at worst be poles, and hence $f$ and $g$ are rational functions (or sections of line bundles) on the CB.

The meromorphic one-form, $\l$, on the elliptic fiber satisfies a differential constraint,
\begin{align}\label{SK2}
\frac{\del\l}{\del u} = \w + d_{X_u} \vf ,
\end{align}
where $\w$ is a holomorphic one-form on the fiber $X_u$, $\vf$ is any meromorphic function, and $d_{X_u}$ is the exterior derivative along the fiber.   $\l$ also satisfies a constraint on the residues at its poles related to the flavor symmetry and the structure of the associated mass deformations; we will describe this constraint later.
The periods of $\l$ along the fiber cycles compute the special coordinates on the CB from which the central charge of the $\cN=2$ algebra and the CB K\"ahler metric can be computed.

The one-form $\l$, however, is not uniquely specified because of the freedom to shift it by total derivatives along the fiber.  Instead, the natural invariant object is the holomorphic 2-form \cite{Donagi:1994,sw2,Donagi:1995cf}
\begin{align}\label{SK4}
\Omega := du\wedge \w = d_u\l - du\wedge d_{X_u}\f ,
\end{align}
defined on the total space of the elliptic fibration.  Its integrals over 2-chains in $X$ with boundary a 1-cycle of $X_u$ compute the central charge at the point $u\in$ CB.  
Because the holomorphic fiber one-form, $\w$, can have arbitrary holomorphic dependence on $u$, the 2-form can be written locally as 
\begin{align}\label{SK5}
\Omega = h(u)\, du\wedge \frac{z dx - 2 x dz}{y},
\end{align}
for some holomorphic $h(u)$ which does not identically vanish.   

$h(u)$ is not unambiguously determined. In fact the change of variables, 
\begin{align}\label{SK6}
z\to\til z:= z, \quad 
x\to\til x:=\a^2 x, \quad \text{and}\quad 
y\to \til y:=\a^3 y,
\end{align}
for arbitrary complex analytic $\a(u)$, preserves the Weierstrass form of the curve without changing its complex structure (i.e., $\t(u)$). 
This implies that the $f$, $g$, and $h$ coefficients in \eqref{SK1} and \eqref{SK5} are ambiguous up to rescalings of the form
\begin{align}\label{SK7}
f\to \til f := \a^4 f, \quad
g\to \til g:=\a^6 g, \quad \text{and}\quad 
h\to \til h := \a\, h.
\end{align}
We will call \eqref{SK7} the {\it analytic gauge freedom} of the SW data.  Thus, at least locally in $u$, we can fix this gauge freedom by setting $h=1$.   

So far, we have described the elliptic fibration in terms of three locally analytic functions $f(u)$, $g(u)$, and $h(u)$ modulo the analytic gauge freedom in terms of a local complex coordinate, $u$, on $\cC$.  We now want to understand the global properties of these functions and how to fix the gauge freedom globally.  In particular, we want to know:  How do they extend to the deformed geometry $\cCb$, i.e., including the singular points?  Also, what is their allowed behavior at metric infinity on $\cC$?  

\subsection{Scaling behaviors at finite points and at infinities}
\label{ssSB}

In this subsection we find explicit Weierstrass forms of the SK cone and cusp geometries described in table \ref{table:eps} and which were derived in appendix \ref{irregular}.

\paragraph{Scaling behaviors at finite points.}

Consider the vicinity of a point on the CB with a metric singularity.  This is a point which is supposed to be at a finite distance (as measured by the metric on the CB) from any point in a small enough neighborhood of it.  Let a local complex coordinate (uniformizing parameter) vanishing at this singularity be $u$.  All the discussion of the last subsection goes through, except now we have no condition that $\Omega$ be regular at $u=0$, i.e., $h(u)$ may have a singularity or a zero there.  But we can use our gauge freedom, \eqref{SK6} and \eqref{SK7}, to set $h=1$ by choosing $\a=h^{-1}$.  We will call this the ``regular gauge".  In this gauge, the 2-form becomes (in a $z=1$ patch of $\P^2_{(1,2,3)}$)
\begin{align}\label{reg-gauge}
\Omega = du \wedge \frac{dx}{y}
\qquad\qquad \text{(regular gauge)}.
\end{align}
Note that by going to regular gauge, we may introduce or modify singularities in $f$ and $g$ at $u=0$.

As long as neither $f$ nor $g$ has an essential singularity at $u=0$, then upon scaling in to $u=0$, the lowest power of $u$ dominates in $f$, $g$, and $h$, so we take, using the rescaling freedom,
\begin{align}\label{scale1}
f \sim u^a, \quad g \sim u^b, \quad h=1,
\end{align}
for integer $a$ and $b$.  Then, assigning mass scaling dimensions $\D(\cdot)$ to $x$, $y$, $z$ and $u$, and using the fact that $\D(\Omega)=1$ (since its periods compute masses) we find from \eqref{SK1} and \eqref{SK5} that
\begin{align}\label{scale2}
2\D(y) &= 3 \D(x) = a \D(u) +\D(x)+4\D(z) = b \D(u) + 6 \D(z)
\nonumber\\
1 &= \D(u) +\D(x) + \D(z) - \D(y)
\end{align} 
from which it follows that
\begin{align}\label{scale3}
\D(u) = \frac{4}{4-a} = \frac{6}{6-b}.
\end{align}
Note that if we had allowed essential singularities or accumulation points of singularities, then there would not exist Laurent expansions of $f$ and $g$ around some point(s) which we could take to be at $u=0$, and there would appear to be no well-defined scaling limit as $u\to0$.

Positive scale dimensions, $\D(u)>0$, imply $a < 4$ and $b < 6$.  (Negative scaling dimensions imply metrics where the singualrity is at metric infinity, so can be discarded as not being part of the moduli space, contrary to our initial assumption.)  The different values of $a \mod4$ and $b\mod6$ give the eight classes of scale-invariant singularities shown in table \ref{kodaira}.  These correspond (as they must) to the list of possible special K\"ahler cone geometries found in the last section and listed in table \ref{table:eps}.  In the two cases --- ($a=0\mod4, b=0\mod6$) and ($a=2\mod4,\ b=3\mod6$) --- where $f$ and $g$ can be simultaneously non-vanishing in the scaling limit, there can be inequivalent subleading corrections to scaling where disc, \eqref{defdisc}, vanishes to higher order (or has a pole of lower order) than either $f^3$ or $g^2$.  These give the non-scale-invariant $I^{(m)}_n$ and $I_n^{*(m)}$ series for $n>0$ also shown in table \ref{kodaira}.  
In table \ref{kodaira} we have also recorded the order of the zero or pole of the discriminant, deg(disc), at the singularity, which will be useful later.  

\begin{table}
\centering
$\begin{array}{|c|l|c|c|c|c|c|}
\hline
\multicolumn{7}{|c|}{\text{\bf Possible Weierstrass form scaling behaviors near singularities of a rank 1 CB}}\\
\hline\hline
\text{name} & \multicolumn{1}{c|}{\text{elliptic curve}} & \ \text{deg(disc)}\ \ &\ \D(u)\ \ & \text{opening angle} & M_0 
& \t_0 \\
\hline
II^{*(m)} &
\parbox[b][0.45cm]{4cm}{$\ y^2=x^3+u^{5-6m}$}             
&10-12m &\frac6{1+6m} &2\pi(\frac16+m) &ST 
& \ e^{i\pi/3}\ \\
III^{*(m)}  &\ y^2=x^3+u^{3-4m}x 
&9-12m &\frac4{1+4m} &2\pi(\frac14+m) &S 
& i\\
IV^{*(m)}  &\ y^2=x^3+u^{4-6m} 
&8-12m &\frac3{1+3m} &2\pi(\frac13+m) &-(ST)^{-1} 
& e^{i\pi/3}\\
I_0^{*(m)} &\ y^2=\prod_{i=1}^3\left(x-e_i(\t)\, u^{1-2m}\right)
&6-12m &\frac2{1+2m} &2\pi(\frac12+m) &-I 
& \t\\
IV^{(m)} &\ y^2=x^3+u^{2-6m} 
&4-12m &\frac3{2+3m} &2\pi(\frac23+m) &-ST 
& e^{i\pi/3}\\
III^{(m)} &\ y^2=x^3+u^{1-4m} x 
&3-12m &\frac4{3+4m} &2\pi(\frac34+m) &S^{-1} 
& i\\
II^{(m)}  &\ y^2=x^3+u^{1-6m} 
&2-12m &\frac6{5+6m} &2\pi(\frac56+m) &(ST)^{-1} 
&e^{i\pi/3}\\
I_0^{(m)} &\ y^2=\prod_{i=1}^3\left(x-e_i(\t)\, u^{-2m}\right)
&0-12m &\frac1{1+m} &2\pi(1+m) &I 
& \t\\[1mm]
\hdashline
I^{*(m)}_n &
\parbox[c][1.1cm]{5cm}{
$\ y^2=x^3+u^{1-2m}x^2$\\ ${}\qquad +\L^{-2n/(1+2m)}u^{n+3-6m}\ \ $}
& 6{+}n{-}12m & \frac2{1+2m} & 0\ \text{(cusp)} & {-T^n} 
& i\infty\\
I^{(m)}_n &
\parbox[c][1.1cm]{5cm}{
$\ y^2=(x-u^{-2m})$\\
${}\qquad \cdot(x^2+\L^{-n/(1+m)}u^{n-4m})\ \ $}  
& 0{+}n{-}12m  & \frac1{1+m} & 0\ \text{(cusp)} & {T^n} 
& i\infty\\[0.5mm]
\hline
\end{array}$
\caption{\label{kodaira} Weierstrass forms of rank 1 special K\"ahler singularities with $\Omega = du \wedge dx/y$, i.e., in regular gauge.  $m\ge0$ and $n>0$ are integers.  The third column gives the order of the zero or pole of the discriminant of the curve at $u=0$.  The last four columns are as in table \ref{table:eps}.  The $I^{*(m)}_n$ and $I^{(m)}_n$ singularities have a parameter $\L$ of mass dimension 1, so are not scale invariant.  Their elliptic curves have, in addition to the singularity at $u=0$, further singularities at $u \sim \L^{\D(u)}$ which are not pertinent: the curves are only meant to capture the scaling behavior for $|u| \ll |\L|^{\D(u)}$.}
\end{table} 

A different way of viewing the curves for the singularities in table \ref{kodaira} is to note that by choosing $\a(u)$ in the analytic gauge freedom \eqref{SK7} appropriately, an $X^{(m)}$ singularity can be described by the ``usual" Kodaira fiber curve given by setting $m=0$ in table \ref{kodaira}, but for which $h(u) = u^m$.  In this alternative gauge the 2-form becomes (in a $z=1$ patch) $\Omega = u^m du \wedge \frac{dx}{y}$, and so has an order-$m$ zero at an $X^{(m)}$ singularity.  For what follows, it will be more convenient to use the regular gauge of \eqref{reg-gauge} which gives the curves shown in table \ref{kodaira}.

\paragraph{Scaling behaviors at infinity.}

Since we are describing CBs as deformations of SCFT CBs, we are necessarily describing $\cN=2$ field theories which are RG flows between UV and IR CFTs (or free theories).  As discussed in detail in \cite{paper1}, such flows are reflected in CB geometries which are deformed in a region close to the UV conformal vacuum, but asymptote to the geometry of the CB of the undeformed UV CFT at large distances.  Thus $\cCb$ will approach $\cCb_0$ asymptotically at the metric infinities, and so will look like a bouquet of cones there.  Each cone, $\cC^0_a$, is parametrized by a specific uniformizing parameter $u_a\in\C^*$. This means that the uniformizing parameter of the deformed geometry $\cCb$ should asymptote to $u_a$ in the vicinity of the $a$th infinity. Furthermore, since by scale invariance $\cC^0_a$ has no metric singularities at large $u_a$, $\cCb$ has also a well-behaved scaling limit as $u_a\to\infty$. This can be turned in a powerful constraint on $f$ and $g$, that is they have to be meromorphic on the \emph{compact} Riemann surface
\begin{align}\label{}
\cCh := \text{$N$-point compactification of}\ \cCb,
\end{align}
by adding a point $\n_a=0$ for $\n_a := 1/u_a$, $a=1,\ldots,N$, for each of the $N$ asymptotic uniformizing parameters. This follows from the fact that at each asymptotic infinity $\cCb\sim\cC^0_a$ and thus $f$ and $g$ can only have an isolated singularity at $\n_a =0$, which corresponds to $u_a\to\infty$. Thus $f$ and $g$ are meromorphic functions (or sections of line bundles) on the compact $\cCh$ Riemann surface, and so only have a finite number of zeros and poles.

(As a check on our reasoning, note that this means that $f$ and $g$ must each have a largest power of $u$ which dominate their large-$u$ expansions in each asymptotic region.  Calling the largest powers of $u$ of $f$ and $g$, $a$ and $b$, respectively, a similar analysis as before gives \eqref{scale3}, where now $\D(u)$ is interpreted as the scaling dimension of $u$ at $\infty$.  Since distances on the CB have mass dimension 1, the distance to infinity is proportional to $\int^\infty |u|^{\D(u)^{-1} - 1} d|u|$, which is infinite iff $\D(u)> 0$.  
This only allows $a<4$ and $b<6$, and the leading scaling behaviors give the same eight series of scale-invariant geometries (asymptotically as $u\to\infty$) as appear in table \ref{kodaira}.  We thus recover precisely the asymptotic scaling forms which we argued are allowed based on the physics of how RG flows of relevant deformations affect the geometry of the CB.)


Note that if we choose to work on the compactified CB, $\cCh$, instead of the affine one, $\cCb$, then the local uniformizing parameter at the infinities of $\cCb$ are the $\n = 1/u$.  Thus the two-form in the gauge given by \eqref{reg-gauge} on $\cCb$ becomes $\Omega \sim \frac{d\n}{\n^2} \wedge \frac{dx}{y}$,
which is \emph{not} in regular gauge on $\cCh$.  A gauge transformation \eqref{SK7} by $\a = \n^2$ brings $\Omega$ to regular gauge on $\cCh$.  This gauge transformation takes the SW curve coefficients to $f \to \til f = \n^8 f$ and $g\to \til g= \n^{12} g$.  By comparison to table \ref{kodaira} we see that this implies that in regular gauge $f$ and $g$ will have the local behaviors near the compactification points given in table \ref{cpctable}.

\begin{table}
\centering
$\begin{array}{|c|l|c|c|c|}
\hline
\multicolumn{5}{|c|}{\text{\bf Scaling behaviors near infinities of a rank 1 CB}}\\
\hline\hline
\text{name} & \multicolumn{1}{c|}{\text{elliptic curve}} 
& \ \text{deg(disc)}\ \ & \ \ \D(\n)\ \ \ & M_\infty \\
\hline
II^{*(m)}_\text{UV} 
&\ \parbox[b][0.45cm]{4cm}{$y^2=\ x^3+\n^{6m+7}$} &14+12m 
&m+\tfrac16
&(ST)^{-1} \\[1mm]
\ III^{*(m)}_\text{UV}\ \ 
&\ y^2=x^3+\n^{4m+5}x &15+12m 
&m+\tfrac14
&S^{-1} \\[1mm]
IV^{*(m)}_\text{UV} 
&\ y^2=x^3+\n^{6m+8} &16+12m 
&m+\tfrac13
&-ST \\[1mm]
I^{*(m)}_{0\,\text{UV}} 
&\ y^2=\prod_{i=1}^3\left(x-e_i(\t)\, \n^{2m+3}\right) &18+12m 
&m+\tfrac12
&-I \\[1mm]
IV^{(m)}_\text{UV} 
&\ y^2=x^3+\n^{6m+10} &20+12m 
&m+\tfrac23
&-(ST)^{-1} \\[1mm]
III^{(m)}_\text{UV} 
&\ y^2=x^3+\n^{4m+7} x &21+12m 
&m+\tfrac34
&S \\[1mm]
II^{(m)}_\text{UV}  
&\ y^2=x^3+\n^{6m+11} &22+12m 
&m+\tfrac56
&ST \\[1mm]
I^{(m)}_{0\,\text{UV}} 
&\ y^2=\prod_{i=1}^3\left(x-e_i(\t)\, \n^{2m+4}\right) \ \ &24+12m
&m+1
&I \\[1mm]
\hline
\end{array}$
\caption{\label{cpctable} Weierstrass forms of rank 1 special K\"ahler affine infinities at $\n=0$ where $\Omega = d\nu \wedge dx/y$, i.e., in regular gauge.  $m\ge0$ and $n>0$ are integers.  The third column gives the order of the zero of the discriminant of the curve at $\n=0$, and the fourth column gives the counter-clockwise $\SL(2,\Z)$ monodromy around $\n=0$.}
\end{table} 

This means that in regular gauge, the $N$ compactified infinities on $\cCh$ are treated differently, since the allowed behaviors of the SW curve in table \ref{cpctable} are different from the allowed behaviors (given in table \ref{kodaira}) at all other points.  In particular, disc must vanish at $\n=0$ at least as $\n^{14}$.

\subsection{Global analysis of SW curve and form}
\label{ssGA}

We now make the following 
\begin{align}\label{smoothass}
\text{{\bf Temporary assumption: }} & \text{the generic deformation of the CB is a}\nonumber\\[-1.2mm]
&\text{smooth complex manifold.}
\end{align}
This implies that $\cCh$ will be a regular Riemann surface, $\cCh_{g,N}$, whose topology is characterized by its genus, $g$, and the number of punctures, $N$.  The punctures are the affine infinite directions and their number equals the number of conical components of the undeformed CB geometry $\cCb^0$.  The physical meaning of this generic smoothness assumption is not clear.   For example, as noted in appendix \ref{appB}, there are singular CBs for which there are no flat deformations enjoying this smoothness property.  However, as we will explain below, the main physically important result remains true even when this additional smoothness assumption is ultimately lifted.

Even if all complex singularities are removed by assumption \eqref{smoothass}, there will still be a co-dimension-1 variety (i.e., a set of points in our rank-1 case) where the \emph{metric} on $\cCh$ is singular or incomplete. Those are the ``standard'' Seiberg-Witten singularities.  In the vicinity of the metric singularities the SW data will asymptote to one of the entries in table \ref{kodaira}, while at metric infinity it will asymptote to one of the entries in table \ref{cpctable}.  With these behaviors fixed at the locus where the SW curve becomes singular and at metric infinity, we then effectively determine the coefficients of the elliptic curve and 2-form by analytic continuation.  
 
Our main result in this subsection will be to show that if and only if there are non-trivial SCFT CB chiral ring relations, then the geometry of its deformed CB at at least one of the metric singularities at finite distance must be of irregular type, i.e., with $m>0$ in table \ref{table:eps}.  Since --- as we will discuss in section \ref{s:dis} --- none of the singularities of irregular type can have the interpretation of IR free field theories, it follows that SCFTs with non-freely-generated rank-1 CBs necessarily flow to exotic fixed point theories.

\paragraph{Patching SW data over smooth CBs.}

To summarize the results of the previous subsection, we can think of the generically deformed CB as a smooth compact Riemann surface, $\cCh_{g,N}$, of some genus $g$ with $N$ marked points.  At the generic point $f$ and $g$ are regular and non-vanishing.   At a finite number of points other than the marked points, where $f$ or $g$ have zeros or poles, there may be singular fibers of any of the types listed in table \ref{kodaira}.  At the marked points the fiber must have a singularity of a type listed in table \ref{cpctable}.

We now want to see how to patch these local conditions together globally over $\cCh_{g,N}$.   The fiber coordinates $[z:x:y] \in \P^1_{(1,2,3)}$ can be thought of as taking values in some holomorphic line bundles, $\cL_z$, $\cL_x$, and $\cL_y$, respectively, over the base $\cCh_{g,N}$.\footnote{In any neighborhood in $\cCh_{g,N}$ with local coordinate $u$, a change of coordinates on $\cL_z$ by $z\to \l(u) z$ for $\l$ a non-vanishing holomorphic function does not change the line bundle $\cL_z$ as a complex variety.  Thus the projective equivalence, $(z,x,y) \sim (\l z, \l^2 x, \l^3 y)$ for $\l\in\C^*$, is a well-defined equivalence on sections of $\cL_z\oplus\cL_x\oplus\cL_y$.}  Likewise, we can think of $f(u)$, $g(u)$, and $h(u)$ appearing in the elliptic curve and the 2-form as sections of some holomorphic line bundles, $\cL_f$, $\cL_g$, and $\cL_h$, over $\cCh_{g,N}$.  The requirements of the existence as a complex space of the total space $X\to \cCh_{g,N}$ of the elliptic fibration over the CB together with the existence of $\Omega$ as a 2-form on $X$ constrain how these various line bundles are related.

The only fiber changes of variable between local patches on $\cCh_{g,N}$ compatible with the $\P^2_{(1,2,3)}$ structure are
$z' = \z_0(u) z$, $x' = \x_0(u) x + \x_1(u) z^2$, and $y' = \up_0(u) y + \up_1(u) xz + \up_2(u) z^3$, for holomorphic $\z_0$, etc.  The $\x_1$, $\up_1$, and $\up_2$ functions can be non-zero only if $\cL_x = \cL_z^2$, $\cL_y = \cL_x \cL_z$, and $\cL_y = \cL_z^3$, respectively, where line bundle tensor products are understood.  However, demanding that the curve $X_u$ be given in Weierstrass form \eqref{SK1} in every local patch on $\cCh_{g,N}$ implies that $\x_1=\up_1=\up_2=0$, since otherwise there would be $xyz$, $yz^3$, or $x^2 z^2$ terms appearing in the curve.  Thus $z$, $x$, and $y$ are indeed sections of independent line bundles with transition functions $\z_0$, $\x_0$, and $\up_0$, respectively.   Define the transition functions of the $\cL_f$ and $\cL_g$ lines bundles to be $f'=\f_0(u) f$, $g'=\g_0(u) g$.  Then matching the forms of the curve in the two patches further implies that $\up_0^2 = \x_0^3 = \f_0\x_0\z_0^4 = \g_0\z_0^6$, or, in terms of line bundles, that $\cL_y^2 = \cL_x^3 = \cL_f\cL_x\cL_z^4 = \cL_g\cL_z^6$.
We can solve these as
\begin{align}\label{hlb1}
\cL_x &= \cL^2 \cL_z^2, &
\cL_y &= \cL^3 \cL_z^3, &
\cL_f &= \cL^4, &
\cL_g &= \cL^6,
\end{align}
for some undetermined line bundles $\cL$ and $\cL_z$.  Note that the definition of the discriminant \eqref{defdisc} implies that disc is a meromorphic section of a discriminant line bundle given by
\begin{align}\label{hlb1a}
\cL_\text{disc} = \cL^{12}
\end{align}
by \eqref{hlb1}.

Let $h' = \eta_0(u) h$ be a transition function of $\cL_h$.  Then
\begin{align}\label{}
\Omega' = h' du' \wedge \frac{z'dx'-2x'dz'}{y'} 
= \frac{\eta_0 \xi_0 \z_0}{\up_0} \frac{du'}{du} \cdot 
h du \wedge \frac{zdx-2xdz}{y} 
= \frac{\eta_0 \xi_0\z_0}{\up_0} \frac{du'}{du} \cdot \Omega.
\end{align}
Since $\Omega$ projects to a globally defined one-form over $\cCh_{g,N}$, it transforms as a section of the canonical bundle, $\cK$, over $\cCh_{g,N}$.  Comparing to the above transition functions then implies that $\cK = \cL_h \cL_x \cL_z \cL_y^{-1}$, or, using \eqref{hlb1},\footnote{The equation \eqref{SK1} defining the elliptic fiber is unaffected if it is multiplied by an overall non-identically-vanishing meromorphic function in each patch.  By choosing this function appropriately in each patch we can trivialize $\cL_z$ if we wish.
This explains why $\cL_z$ cancels in the determination of the $f$, $g$, and $h$ bundles.}
\begin{align}\label{hlb3}
\cL_h = \cK\cL.
\end{align}

Finally, we have the analytic gauge freedom, \eqref{SK7}, which allows us to multiply $h$ by an arbitrary meromorphic function $\a(u)$ in each patch.  Thus, by choosing this function appropriately in each patch we can change the transition functions of $\cL_h$ to that of any other line bundle.  In particular, the regular gauge \eqref{reg-gauge} which sets $h=1$ in each patch (including patches including the $N$ marked points at metric infinity) trivializes $\cL_h$, and thus we have from \eqref{hlb3}, \eqref{hlb1}, and \eqref{hlb1a} that
\begin{align}\label{fgdiscK}
\cL_f &= \cK^{-4}, \qquad \cL_g = \cK^{-6}, 
\qquad \cL_\text{disc} = \cK^{-12}
\qquad\qquad
\text{(regular gauge)}.
\end{align}

Since deg$(\cK)=2g-2$, this implies that deg$(\cL_\text{disc}) = 24(1-g)$.  But for each affine infinity, the discriminant has a zero of order at least 14, therefore the divisor of the discriminant at finite points has degree  deg$(\cL_\text{disc}|_{\cCb}) \le 2(12-12g-7N)$.  In particular, unless $g=0$ and $N=1$ the discriminant must have poles at finite points on $\cCh_{g,N}$.   From table \ref{kodaira} we see that only irregular ($m>0$) singularities contribute poles to the discriminant.  Therefore we conclude that
\begin{align}\label{irregnp}
\cCh_{g,N} & \ \text{necessarily has at least one irregular singularity unless $g=0$ and $N=1$.}
\end{align}

If $g=0$ and $N=1$ and the CB is deformed to be smooth as a complex manifold (so $\cCb \simeq \C$), then deg$(\cL_\text{disc}|_{\cCb}) \le 10$.  This inequality can be satisfied with just regular ($m=0$) singularities.   These are precisely the ``planar" regular rank-1 CBs which were systematically constructed and analysed in \cite{paper1, paper2, allm1602, am1604, paper3}.

Furthermore, when $g=0$ and $N=1$ \emph{only} regular singularities can appear on $\cCb$.  The reason is that if $\cCb\simeq\C$ (which is the content of assuming $g=0$ and $N=1$), then the chiral ring of the undeformed $\cCb_0$ must have been freely generated. In fact the genus of a SCFT CB $\cCb_0$ with a complex singularity at the origin is strictly less than that of its smooth deformation, $\cCb$.\footnote{This can be seen as follows. All (reduced) Riemann surfaces can be embedded in $\P^M$ by polynomial equations for large enough $M$, and each is birationally equivalent to its plane  projection.  The genus of a singular plane curve is given by the genus formula, $g = \frac12 d(d-1) - \sum_P \d_P$ where the positive integers $d$ and $\d_P$ are the degree of the plane curve and the delta invariants of the singular points $P$; see, e.g., section I.3.4 of \cite{GLS07}.}  Thus if $\cCb \sim \C$ (so has $g=0$), then we must have $\cCb_0 \sim \C$ as well, and so its coordinate ring is freely generated.  It then follows that the uniformizing parameter $u$ is the vev of some CB chiral operator in the SCFT, and so $\D(u) \ge1$ by unitarity.  This condition excludes all the irregular SK singularities from appearing.

The conclusion \eqref{irregnp} is in some sense the easiest implication of \eqref{fgdiscK}, following only from the degree of the canonical bundle on a genus-$g$ Riemann surface.  In principle much more detailed implications relating the complex structure of $\cCh_{g,N}$ to the nature of the metric singularities follow from the Riemann-Roch theorem.

\paragraph{Closedness of the set of theories with non-freely-generated CB chiral rings under RG flows.}

Irregular singularities, from table \ref{table:eps}, all have $\D(u)<1$ where $u$ is the local uniformizing parameter with the singularity at $u=0$.  Thus, as discussed around \eqref{unitarity}, such theories can only be consistent with unitarity if their local uniformizing parameter is not the vacuum expectation value of any field in the SCFT.  This can only happen if there are non-trivial relations among the fields of the CB chiral ring at $u=0$.   Thus: 
\begin{align}\label{result}
&\text{Rank-1 unitary SCFTs with non-freely-generated CB chiral rings flow to}\nonumber\\[-1.2mm]
&\text{at least one IR fixed point with a non-freely-generated CB chiral ring.}
\end{align}
Note that this weaker formulation of \eqref{irregnp} holds in more generality:  even if the assumption \eqref{smoothass} of the generic smoothness of the deformed CB does not hold, \eqref{result} is valid.  The reason is simply that any points where the deformed CB is not a smooth complex manifold are, as we discussed in section \ref{s:sir1}, points where there are non-trivial CB chiral ring relations.

\eqref{result} is the main result of this paper.  It shows that (rank-1) SCFTs with non-freely-generated CB chiral rings must form a distinct subset under RG flows since they can only flow into one another.

\subsection{Construction of examples of non-planar SK geometries}
\label{ssCC}

The deformation $\cCb$ of the SCFT CB $\cCb^0$ is subject to further conditions coming from demanding that a special K\"ahler (SK) structure on $\cC$ is preserved.  This SK condition ties together the deformation of the CB as a complex variety and the deformation of the elliptic fibration in two overlapping ways.  
\begin{itemize}
\item[$(i)$]
The structure of the CB as a complex variety is closely tied to that of the elliptic fibration (the SW curve) for non-generic values of the deformation parameters.  This is because at any complex singularity of $\cCb$ the local coordinate ring is not freely generated, as discussed in section \ref{2.1}.  But an $\cN=2$ $\U(1)$ gauge theory (with no massless matter) has a freely-generated chiral ring, so the IR theory at any complex singularity must be a theory where  extra massless BPS states appear.  These correspond to metric singularities, i.e., to vacua where degenerations of the elliptic fibers occur.  Thus the CB must depend on the deformation parameters in such a way that wherever it develops a complex singularity the SW curve also degenerates.
\item[$(ii)$]
The periods of the Seiberg-Witten form (computing the central charge) must vary linearly with the dimension-1 mass parameters, $\bm$, which can be taken to be in a complexified Cartan subalgebra of the flavor algebra, and which enter polynomially in the deformation parameters of the CB and the elliptic fibers.  This is usually expressed \cite{sw2} as the condition that the residues of the SW 1-form satisfy $\text{Res}(\l) \in \{ \ba(\bm)\}$, where $\ba$ is in the flavor algebra root lattice (so $\ba$ acts linearly on the $\bm$).  In terms of the invariant 2-form on the total space $X$, this becomes the condition that the periods of $\Omega$ on closed 2-cycles of $X$ compute $\ba(\bm)$. This condition it is also discussed extensively in \cite{paper2,Minahan:1996fg,Minahan:1996cj}.  
\end{itemize}
$(ii)$ is a stringent condition, and is not wholly independent of condition $(i)$.  

Although we are not able to solve these consistency conditions in  generality, we will now show how to construct many examples of non-planar CB geometries as multi-sheeted covers of planar SK CB solutions which do satisfy them. This shows that the requirement of a SK structure is not by itself an obstruction to the existence of CBs describing deformations of SCFTs with non-freely-generated CB chiral rings.

Say the uniformizing parameter of the planar CB solution is $\f\in\C$, so the SW curve and form are
\begin{align}\label{planarSW}
y^2 = x^3 + f(\f,\bm) x + g(\f,\bm), \qquad
\Omega = d\f \wedge \frac{dx}{y}, \qquad \f\in \C,
\end{align}
for some $f$ and $g$ polynomial in $\f$ and the linear masses $\bm$.  Let's suppose (just for definiteness for illustrative purposes) that the UV singularity at $\f=0$ when $\bm=0$ is, say, a $IV^{*(0)}$ singularity and therefore $\f$ has scaling dimension $\D(\f)=3$.  Furthermore, suppose that for generic $\bm$ the UV singularity is deformed into a collection of two $I_1^{(0)}$ singularities and two $I_3^{(0)}$ singularities.\footnote{In the language of \cite{paper1, paper2, allm1602, paper3, am1604}, this would be a geometry with deformation pattern $IV^{*(0)} \to \{(I_1^{(0)})^2, (I_3^{(0)})^2\}$.  This particular geometry did not appear in those papers since it fails to obey the Dirac quantization condition on the CB.  However in the present context, it is simply an intermediate step in constructing a non-planar SK CB geometry, and the Dirac quantization constraint does not apply.  It is was shown in \cite{aw10} that this deformation can actually be realized by turning on mass deformations appearing as invariant polyonimials of degree 2 and 6.}  The positions of these IR singularities are given by the zeros of the discriminant.  Since there are two different types of IR singularities, they cannot mix under monodromies in the space of mass parameters, and therefore the discriminant must factorize as (see \cite{paper1})
\begin{align}\label{DisCB}
\text{disc} = P_2(\f,\bm) \cdot Q_2(\f,\bm)^3
\end{align}
where $P_2$ and $Q_2$ are weighted-homogeneous polynomials quadratic in $\f$.  The two zeros of $P_2$ give the positions in the $\f$-plane of the $I_1^{(0)}$ singularities and the two zeros of $Q_2$ give the positions in the $\f$-plane of the $I_3^{(0)}$ singularities.

Now let us simply consider the non-planar CB given by the Riemann surface
\begin{align}\label{CBRS}
\cCb: \quad \psi^n = Q_2(\f,\bm), \qquad (\f,\psi)\in\C^2, 
\qquad \Z\ni n > 1.
\end{align}
We have left $n$ unspecified just to illustrate a variety of different non-planar geometries that we can construct in this way.
\eqref{CBRS} describes an $n$-fold cover of the $\f$-plane branched over two points (or, by completing the square in $Q_2$, as a 2-fold cover of the $\psi$-plane branched over $n$ points).  When $n$ is odd, this describes a hyperelliptic genus $g=(n-1)/2$ Riemann surface with a single puncture corresponding to the asymptotic infinity, while for $n$ even it describes a hyperelliptic genus $g=(n-2)/2$ Riemann surface with a two punctures corresponding to two asymptotic infinities.  

The CB geometry defined by \eqref{CBRS} automatically satisfies condition $(i)$ and $(ii)$. First of all the locus of complex singularities (see subsection \ref{2.1}), $\Sigma$, are the points satisfying
\beq
\Sigma\ :\quad
\left\{
\psi = Q_2(\f,\bm) = \frac{\del Q_2(\f,\bm)}{\del \f} = 0
\right\} .
\eeq 
This ensures that the complex singularities are a subset of the vanishing locus of the discriminant \eqref{DisCB}, and thus ensures that the elliptic fiber degenerates at all the complex singularities. Furthermore, since the residues of the SW form for the planar geometry were linear functions of the $\bm$, it is easy to see that this remains true of the theory on the non-planar CB with the same SW form and thus $(ii)$ follows. 

The geometry defined by \eqref{CBRS} has a tight and interesting structure. At generic $\bm$ the two $\f$-plane branch points coincide with the positions of the two $I^{(0)}_3$ singularities in the planar theory.  However on the CB given in \eqref{CBRS} the geometries of these  singularities are changed.  The reason is that at the branch points $\f$ is no longer a uniformizing parameter of the CB, but instead $u \sim (\f-\f_*)^{1/n}$ is (for $\f_*$ a root of $Q_2$).  This means that in these variables the holomorphic 2-form of \eqref{planarSW} has the form $\Omega \sim u^{n-1} du \wedge \frac{dx}{y}$, and, as we remarked in section \ref{ssSB}, this is characteristic of an \emph{irregular} singularity with $m=n-1$.  Furthermore, the $\SL(2,\Z)$ monodromy in the $\f$-plane around an $I^{(0)}_3$ singularity is in the $T^3$ conjugacy class, so the monodromy on the Riemann surface \eqref{CBRS} where this is a branch point with ramification index $n$ has conjugacy class $(T^3)^n = T^{3n}$.   We thus see that with the non-planar CB, the branch points correspond to $I^{(n-1)}_{3n}$ singularities in the geometry.  Meanwhile, the two original $I^{(0)}_1$ singularities at the zeros of $P_2$ do not coincide with branch points and so their geometry remains unchanged; however their multiplicity is increased by a factor of $n$ since there are now $n$ copies of each of these points --- one for each sheet of the Riemann surface \eqref{CBRS}.  In summary, at generic $\bm$, the CB is (as a complex manifold) the smooth Riemann surface described in the previous paragraph with two irregular $I^{(n-1)}_{3n}$ metric singularities and $2n$ regular $I^{(0)}_1$ metric singularities.

In particular, in the conformal limit, where $\bm=0$, \eqref{CBRS} becomes the SCFT CB chiral ring relation $\psi^n=\f^2$.  In the case of $n$ odd, this describes a conical CB geometry with uniformizing parameter $u = \vev{\psi}^{1/2} = \vev{\f}^{1/n}$ which therefore has dimension $\D(u)=3/n$.  Then, by comparison to table \ref{table:eps}, we see that this scale invariant CB geometry is that of an SK cone of type
\begin{align}\label{}
\begin{cases}
IV^{*(m)} \ \text{with}\ m=\frac{n-1}{3} 
& \text{if}\ n = 1 \text{ (mod 3)} \\
IV^{(m)} \ \text{with}\ m=\frac{n-2}{3} 
& \text{if}\ n = 2 \text{ (mod 3)} \\
I_0^{(m)} \ \text{with}\ m=\frac{n-3}{3} 
& \text{if}\ n = 0 \text{ (mod 3)}
\end{cases}
\qquad\qquad \text{($n$ odd).}
\end{align}
If $n$ is even, the CB is a bouquet of two cones each with uniformizing parameter $u=\vev\psi$ so of dimension $\D(u)=6/n$ and so describing an SK cone of type
\begin{align}\label{}
\begin{cases}
IV^{*(m)} \ \text{with}\ m=\frac{n-2}{6} 
& \text{if}\ n = 2 \text{ (mod 3)} \\
IV^{(m)} \ \text{with}\ m=\frac{n-4}{6} 
& \text{if}\ n = 1 \text{ (mod 3)} \\
I_0^{(m)} \ \text{with}\ m=\frac{n-6}{6} 
& \text{if}\ n = 0 \text{ (mod 3)}
\end{cases}
\qquad\qquad \text{($n$ even).}
\end{align}

With this example we have shown how to construct (infinitely many) non-planar CB SK geometries simply by taking multi-sheeted covers of a planar CB SK geometry.  There are clearly many other examples that can be constructed along these lines.  This construction can only describe rank 1 CBs which are given as plane curves, and so whose corresponding SCFT CB chiral rings have two generators and a single relation.  Presumably SK geometries for CBs given as deformations of more complicated chiral rings also exist, but we do not know of a systematic way to construct them.

\section{Discussion: physical interpretation of irregular singularities}
\label{s:dis}

Our main conclusions of the last two sections are that:
\begin{itemize}
\item irregular CB singularities are consistent with unitarity only if the associated field theory at the singularity has non-trivial chiral ring relations,
\item unitary SCFTs with non-trivial chiral ring relations must generically flow to at least one ``frozen" IR singularity which is either an irregular singularity or a regular singularity but which is at a complex manifold singularity of the CB,
\item and in either of the above cases the theory at the IR singularity will have non-trivial chiral ring relations.
\end{itemize}
These conclusions do not rule out the existence of rank-1 SCFTs with non-freely-generated CB chiral rings.  It is logically possible that such SCFTs could flow to as-yet-unknown irregular ``frozen" (i.e., lacking any relevant $\cN=2$-preserving deformation) SCFTs.  But we do not know of any independent evidence for (or against) the existence of such novel IR fixed point theories.

One case where we have special control over SCFT RG flows is when all the IR fixed points are weakly coupled gauge theories.  Indeed, 
a large subset of rank-1 SCFTs with freely-generated CB chiral rings generically flow to IR free field theories, and their analysis formed the basis of the classification given in \cite{paper1, paper2, allm1602, am1604, paper3}.  A natural question, therefore, is whether there is a similar subset of SCFTs with non-freely-generated CB chiral rings admitting a weakly-coupled low energy interpretation.  In particular, our results allow the possibility that such a rank-1 SCFT could flow to only $I^{(m)}_n$ and $I^{*(m)}_n$ singularities with $n>0$ and at least one $m>0$.  The low energy $\U(1)$ gauge coupling at these singularities are $\t_0=i\infty$, i.e., they flow to zero coupling.  

Nevertheless, we will now argue that the IR singularities with $m>0$ do not have any familiar interpretation in terms weakly coupled field theory, and thus must involve some ``exotic" physics (if in fact they exist at all).   Furthermore, we propose that the exotic physics is that weakly gauging flavor symmetries of SCFTs with CB chiral ring relations must involve quantum modifications of those relations.

Firstly, the $I^{(m)}_n$ and $I^{*(m)}_n$ singularities with $m>0$ cannot be IR free theories simply because the IR free field theories with rank-1 CBs are known: they are $\U(1)$ or $\SU(2)$ gauge theories with sufficient numbers of massless matter hypermultiplets, and, as such, the dimensions of their CB parameters are 1 or 2, respectively.  Thus they can correspond only to the (regular) $I^{(0)}_n$ or $I^{*(0)}_n$ singularities in table \ref{table:eps}.  

But the fact that the $\U(1)$ coupling on the CB goes to zero at the singularity suggests that the low energy field theory description of the theory at the singularity must be that of a vector multiplet weakly gauging a rank-1 --- i.e., $\U(1)$ or $\SU(2)$ --- subalgebra of the flavor symmetry of some ``matter theory".  By a matter theory, we just mean some scale-invariant rank-0 theory --- i.e., one with no CB.  The only rank-0 examples we know of are free massless hypermultiplets.  Nevertheless, suppose there exists interacting rank-0 $\cN=2$ SCFTs with flavor symmetries which contain $\U(1)$ or $\SU(2)$ subalgebras with flavor central charge $\k=n$ or $\k=8+n$, respectively.  Coupling to a $\U(1)$ or $\SU(2)$ vector multiplet by weak gauging then gives an IR free rank-1 CB with beta function $\propto n$, and so a CB with a $T^n$ or $-T^n$ EM duality monodromy, respectively.  These can thus only be described by the $I^{(m)}_n$ and $I^{*(m)}_n$ geometries.

This description seems to imply that that the CB parameter will be given by the gauge-invariant vevs of the vector multiplet adjoint scalar, $\f$, namely $u = \vev\f$ or $u=\vev{\Tr \f^2}$, respectively.   If that were the case, then $\D(u)=1$ or $2$, and only the regular $I^{(0)}_n$ and $I^{*(0)}_n$ geometries would be realized.   However, a possible way to avoid this conclusion, and so to realize the irregular $I^{(m)}_n$ and $I^{*(m)}_n$ geometries is as follows.  

The irregular geometries have CB parameters with scaling dimensions less than one, so violate unitarity if they are vevs of SCFT fields.  Now suppose that there is a CB chiral ring relation between the vector multiplet scalar,\footnote{Here we are assuming that the vector multiplet gauges a $\U(1)$ symmetry, so $\f$ is gauge invariant and has dimension $\D(\f)=1$.  Similar examples could be constructed for $\SU(2)$ vector multiplets with $\Tr(\f^2)$ in place of $\f$.} $\f$, and some other $\cE_r$-type chiral field, $\psi$, in the rank-0 theory.  In the limit in which we turn off the IR free $\U(1)$ gauge coupling, $\f$ should become free, and so not enter in any chiral ring relations.   Thus, as we take the $\U(1)$ strong coupling scale $\L\to\infty$, all terms in the chiral ring relations involving $\f$ should vanish.  But since the rank-0 theory has no CB, $\psi$ must not be allowed to get a vev.  One way of enforcing this is to suppose that $\psi$ is nilpotent, so $\psi^b=0$ for some integer $b>1$ when $\L=\infty$.  A natural way to satisfy these requirements is if the chiral ring relation of the weakly gauged rank-0 theory plus $\U(1)$ vector multiplet satisfies a quantum-corrected chiral ring relation like
\begin{align}\label{qcorrcr}
\psi^b = \L^{-n} \f^a
\end{align}
for some relatively prime\footnote{If $a$ and $b$ were not relatively prime, then there would be lcd$(a,b)$ CB components emanating from the singularity.} integers $a>b>1$ and where $n$ is the beta function coefficient of the $\U(1)$ coupling.   Then the CB uniformizing parameter is\footnote{The reason $u$ is $\vev{\f}^{1/b}$ and not, say, $\L^c \vev{\f}^{1/b}$ for some non-zero $c$ is because in this non-scale-invariant situation we are actually interested in the scaling dimension of the CB parameter in the limit as we approach the singularity.  Equivalently, we want a uniformizing parameter which remains finite and non-zero at fixed $\f$ while $\L\to\infty$, and only $\vev{\f}^{1/b}$ does this.  Note that in this limit \eqref{qcorrcr} implies that $\vev{\psi}\to0$, which is consistent with being at the conformal point of the rank-0 theory.} $u = \vev{\f}^{1/b} = \L^{n/ab}\vev{\psi}^{1/a}$ and so has dimension $\D(u)=\frac1b$.  This has the right dimension to be a uniformizing parameter for an $I^{(m)}_n$ singularity if $b=m+1$, and satisfies the unitarity constraint $\D(\psi)>1$ if $a>n+b$.

This gives a coherent field theory picture of how an $I^{(m)}_n$ or $I^{*(m)}_n$ irregular singularity could arise from a field theory, but only if weakly gauging a flavor symmetry could give rise to a quantum-corrected chiral ring relation as in \eqref{qcorrcr}, and if rank-0 SCFTs with nilpotent CB-type operators exist.   We'd now like to argue that, if one admits the possibility of SCFTs with non-trivial CB chiral ring relations, then both the above ingredients are in some sense ``natural" consequences.

For example, consider a hypothetical rank-2 SCFT with CB chiral ring generated by three fields, $\f$, $\psi$, and $\chi$, but satisfying a relation
\begin{align}\label{qcorrcr2}
\chi \psi^b = \f^a , \qquad a>b>1.
\end{align}
Thus this CB has a complex singularity along $\f=\psi=0$ for all $\chi$.   If the monodromy around this singularity is of $T^n$ type\footnote{More precisely, the monodromy is in $\Sp(4,\Z)$, but we are assuming it is $T^n \in \SL(2,\Z)\subset \Sp(4,\Z)$ where the $\SL(2,\Z)$ is the subgroup appropriate to the $\U(1) \subset \U(1)^2$ which is ``transverse" to the $\f=\psi=0$ singularity.} then by tuning (or ``flowing to") $\chi =\L^n$ and decoupling the low energy $\U(1)$ which does not participate in the $T^n$ monodromy, we have an effective rank-1 theory with quantum-corrected chiral ring relation as in \eqref{qcorrcr} and an $I^{(m)}_n$ type singularity.

In any case, all the above arguments are merely suggestive, and it would obviously be desirable to have more definitive arguments for or against the existence of QFTs with $I^{(m)}_n$ or $I^{*(m)}_n$ irregular singularities.

\paragraph{Some further open questions.}

We conclude with some obvious open questions.
\begin{itemize}
\item
A different way of approaching the consistency of the irregular SK singularities is to ask whether their low energy effective actions are self-consistent.  In particular, the geometry of the irregular singularities is characterized by the fact that they have curvature singularities at their ``tips" which are negative,\footnote{In the case of the cusp-like irregular $I^{(m)}_n$ or $I^{*(m)}_n$ singularities, the curvature has to be regularized since the singular point contributes positive infinite total curvature while the integral of the curvature over any neighborhood of regular points contributes negative infinite total curvature.} reflected in the fact that they have opening angles greater than $2\pi$.  Is there some instability of a low energy theory with a moduli space geometry of this sort, perhaps along the lines of those discussed in \cite{Adams:2006sv}?  (Note that negative moduli space curvature by itself is not an indication of instability, since all special Kahler geometries have non-positive scalar curvatures at regular points \cite{Freed:1997dp}.)
\item
Another way to test the existence of theories with irregular CB geometries is to look for them using numerical bootstrap methods, as was mentioned in section \ref{s2.3} above.   In particular, one would have to look for numerical evidence for a solution to the $\cN=2$ crossing relations with CB chiral ring relations, since the bootstrap currently cannot probe the CB geometry directly.
\item
Finally, can the arguments we have presented here be extended to higher-rank CBs?  The assumption that the compactified CB is a quasi-projective variety which is a flat deformation of a given UV singularity, together with a higher-dimensional generalization of the Reimann-Roch theorem, plausibly allow us to derive a similar constraint on the degree of the discriminant bundle similar to that found above eqn.\ \eqref{irregnp}.   What is less clear is how the relation between the order of vanishing of the discriminant and the irregularity of a CB metric singularity generalizes to higher rank.  This latter depends not just on the structure of the CB as a complex manifold but more particularly on its special K\"ahler structure.
\end{itemize}

\acknowledgments

It is a pleasure to thank J. Distler, T. Dumitrescu, D. Kulkarni, M. Lemos, M. Lotito, D. Morrison, Y. Tachikawa, and D. Xie for helpful comments and discussions.  This work was supported in part by DOE grant DE-SC0011784.  MM was also partially supported by NSF grant PHY-1151392.





\begin{appendix}

\section{Geometries of irregular CB singularities}
\label{irregular}

We show here how to find all 1-dimensional special K\"ahler (SK) manifolds, $\cC$, with a complex homothety.  This gives the derivation of the results reported in table \ref{table:eps}.  We start with a quick review of the basics of rank-1 SK geometry and then specialize to the systematic study of the scale invariant case.

A rank-1 SK geometry is specified by a choice of ``special coordinate" which is a holomorphic section of a rank-2 vector bundle with structure group $\SL(2,\Z)\subset\GL(2,\C)$
\begin{align}\label{CBscs}
\s := \bpmat a_D\\ a \epmat.
\end{align}
This means that $\s$ is holomorphic in the CB parameter $u$ but may suffer an $\SL(2,\Z)$ monodromy transformation along loops encircling singularities.  Monodromies thus provide a representation of $\pi_1(\cC)$ in $\SL(2,\Z)$, which is interpreted as the electric-magnetic (EM) duality group of the low energy $\U(1)$ gauge theory on the CB.  

The central charge of the $\cN=2$ supersymmetry algebra in a sector with EM charges $z:=(\bsmat p\\q\esmat)$ is given by $Z_z=z^T \s + \text{mass terms}$, and computes the BPS mass bound on states in that sector.  Thus the special coordinates $\s$ have mass dimension 1.

The K\"ahler metric on $\cC$ is then given in terms of a K\"ahler potential $K = \Im(\bar{a}\, a_D)$ in the usual way, 
\begin{align}\label{CBmetric}
ds^2 = g_{u\ub}du d\ub =\del_u \delb_\ub K dud\ub ,
\end{align}
where $u$ is a complex coordinate on $\cC$.  This metric is required to be positive definite, which is equivalent to requiring that $\Im\t >0$ where $\t :=(da_D/du)/(da/du)$ is the low energy EM coupling on the CB.

For higher-rank CBs, there are other requirements on the special coordinates.  Also, even for rank-1 CBs, there are further restrictions on how the special coordinates can depend on the complex mass parameters \cite{sw2} which enter as certain complex deformation parameters of $\cC$, and are related to the (unspecified) mass terms that appeared in the expression for the central charge above.  We describe these constraints further in section \ref{ssCC}.

Now that we have reviewed the definition of rank-1 special K\"ahler geometry, and in particular of the interconnected role played by the metric, the special coordinates and monodromies, we are ready to study scale invariant special K\"ahler geometries.  Scale invariance plus invariance under $U(1)_R$ rotations of the metric on $\cC$ under $u \to \l u$ means that $u\to \l u$ acts as a complex homothety on $\cC$.  This implies from \eqref{CBmetric} that
\begin{align}
\s(\l u) = \l^\e \s(u)
\quad\text{for some}\quad \e\in\C.
\end{align}
Thus
\begin{align}\label{sigep}
\s(u)= u^\e  \bpmat \t_0 \\ 1 \epmat,
\quad\text{for some}\quad \t_0\in\C.
\end{align}
Above we have chosen an overall complex constant factor to set $a(u) = u^\e$.  The metric on $\cC$ is then
\begin{align}\label{met1}
ds^2 = (\Im\t_0)|\e u^{\e -1}|^2 du d\bar{u}.
\end{align}
Positivity of the metric implies that $\Im\t_0>0$.
Writing $u = \r e^{i\f}$ and $\e = \e_1 + i\e_2$, we have $ds^2 = (\Im\t_0) |\e|^2 \r^{2(\e_1-1)}e^{-2\f\e_2}(d\r^2 + \r^2 d\f^2)$, which is periodic only if $\e_2 = 0$; that is, only if $\e$ is real.  The radial distance to the origin is then $\int_0 ds \sim \int_0\r^{\e-1}d\rho$, and is only finite if $\e > 0$.  Physically, since the mass dimension of $\s$ is 1, and $\s \sim u^\e$, we have 
\begin{align}\label{}
\D(u) = 1/\e 
\end{align}
where $\D(\cdot)$ denotes the mass dimension.  The metric \eqref{met1} then implies that $\cC^0$ is a flat cone with tip at $u=0$ and opening angle $2\pi/\D(u)$.   Note that for $\D(u) <1$ the cone has negative curvature at its tip (it has an opening angle greater than $2\pi$) while for $\D(u)>1$ is has postive curvature there (it has an opening angle less than $2\pi$).

\begin{table}
\centering
$\begin{array}{|c|c|c|}
\hline
\ \Tr M_0\ \,&\ \text{conjugacy classes}\ \,&\ \text{eigenvalues}\ \m\ \, \\ \hline
2      & T^n                             & +1                   \\ 
1      & ST, (ST)^{-1}              & e^{\pm i\pi/3}  \\ 
0      & \pm S                          & e^{\pm i\pi/2}  \\ 
-1     & -ST, (-ST)^{-1}           & \ e^{\pm 2i\pi/3}\ \,\\ 
-2     & -T^n                           & -1                      \\ \hline
\end{array}$
\caption{Representatives of all elliptic and parabolic conjugacy classes of $\SL(2,\Z)$. $n$ is an arbitrary integer, and $T := {1\ 1\choose0\ 1}$, $S := {0\ -1\choose1\ \ 0}$ are generators of $\SL(2,\Z)$ satisfying $S^2=(ST)^3=-I$.\label{table:conjclass}
}
\end{table}

The geometry is only singular at $u=0$ and thus $\pi_1(\cC) = \Z$ which can be taken to be generated by $\g$, a path that circles once around $u=0$ counterclockwise. Thus there is only a single non-trivial monodromy, $M_0 \in \SL(2,\Z)$, corresponding to analytic continuation along $\g$.  We can thus describe the special coordinates $\s$ \eqref{CBscs} as holomorphic on the cut $u$-plane with cut chosen (arbitrarily) on the negative imaginary $u$-axis and satisfies $\s(0^- {+} iy) = M_0 \s(0^+ {+} iy)$ for all $y<0$.  That is, $\s$ must ``jump'' by the linear action of $M_0$ across the cut.  As we follow the path $\g$, $\s \to M_0 \s$, and so from \eqref{sigep}
\begin{align}\label{}
e^{2\pi i \e} \bpmat \t_0\\1 \epmat = 
M_0 \bpmat \t_0 \\ 1 \epmat .
\end{align}
$M_0$ must therefore have an eigenvalue, $\m:=e^{2\pi i\e}$, with $|\m| = 1$.  The characteristic equation of $M_0$ is $\m^2 - (\Tr M_0)\m +1 = 0$, since $\det M_0=1$.  It follows that $M_0$ has an eigenvalue of unit norm if and only if $(\Tr M_0)^2 \le 4$.  Representatives of all $\SL(2,\Z)$ conjugacy classes with this property are listed in table \ref{table:conjclass}.    

Since $\m = e^{2\pi i \e}$ we have $\e=\e_0 + m$ for some $\e_0$ and any $m\in\Z$.  Since also $0 < \e =1/\D(u)$, choosing $\e_0$ to be the smallest positive solution for $\e$, then we find an infinite tower of solutions for each value of $\m$ in table \ref{table:conjclass} labelled by $m\ge0$.  This gives the list of allowed values of $\D(u)$, the cone opening angle $2\pi/\D(u)$, the EM duality monodromy $M_0$, and $\t_0$, shown in table \ref{table:eps}.  The entries with $m=0$ all satisfy $\D(u)\ge1$, while the entries with $m\ge1$ all have $\D(u)<1$.  

Note that an $I^{(0)}_0$ ``singularity" has opening angle $2\pi$, so has no metric singularity:  it describes the local scaling behavior of generic, metrically smooth points on the CB.

Finally notice that for the last two rows in table \ref{table:eps} there is actually no scale-invariant solution for $\s=\left(\bsmat a_D \\ a \esmat\right)$ since $\t_0 = i\infty$.  So we must look for solutions by including the leading corrections to scaling.  Since these asymptotically scale-invariant geometries will play an important role in section \ref{s:ru}, we briefly describe them here.  For example, for the $T^n$ monodromy, expanding the special coordinates as
\begin{align}\label{}
\s = \bpmat a_D\\ a\epmat &= u^{m+1} 
\bpmat \frac{n}{2\pi i}\ln\left(\frac{u}{\L}\right) \\
1 \epmat,
\end{align}
where $\L$ is an arbitrary mass scale gives a solution to $\s(e^{2\pi i}u) = T^n \s(u)$.  For this solution the metric is $ds^2 = \del_u \del_{\bar{u}} K du d\bar{u}$, with $K = \Im(\bar{a}\, a_D) = -\frac{n}{4\pi} |u|^{2(m+1)} \ln\left(\frac{u\bar{u}}{\L^2}\right)$, so the final solution for the metric is
\begin{align}
ds^2 = -\frac{n}{4\pi} |u|^{2m} \left\{
(m+1)^2 \text{ln}\left(\frac{u\bar{u}}{\L^2}\right) 
+2 (m+1) \right\} du d\bar{u}.
\end{align}
Note that as $|u|\to 0$, ln$(u\bar{u}) \to -\infty$, so the metric is positive-definite in the vicinity of $u=0$ only for $n>0$.  This metric describes a cusp at $u=0$ with finite radial distance to the tip of the cusp.  Thus only the $T^n$, $n \in \Z^+$ monodromies give physical singularities.  
A similar story goes for the $-T^n$ monodromies, which also give positive definite metrics in the vicinity of the cusp only for $n\in \Z^+$.

\section{Examples of rank-1 chiral rings}
\label{appA}

\paragraph{Example 1: Freely-generated CB chiral ring.}  

The CB chiral ring of this simplest example has a single generating field, $\f$, and no relations.  Thus the CB is simply the $u=\vev\f$ complex plane.  In this case the metric singularity at the origin (the tip of the conical geometry) is not a singularity of the complex structure.  Since the uniformizing parameter, $u$, is the vev of a field, unitarity and scale invariance then restricts the CB geometry to be one of the regular $m=0$ cones in the first eight rows of table \ref{table:eps}.  These then limit the allowed dimensions to the set $\D(u) \in \{ 6, 4, 3, 2, \frac32, \frac43, \frac65, 1\}$.  The distinct special K\"ahler geometries whose scale-invariant limits belong to this class were classified in \cite{paper1, paper2, allm1602, am1604, paper3}.  (In the last case, $\D(u)=1$, the corresponding field, $\f$, is free, and the CB ``cone" is flat --- i.e., the euclidean plane.)  

\paragraph{Example 2: One-component hypersurface CB.}  

The next simplest example is an irreducible chiral ring with two generators, $\{\f_1,\f_2\}$, and a single relation.  If the ring is irreducible and reduced (without nilpotents), then homogeneity implies its ideal must be generated by a relation of the form
\begin{align}\label{CRex2}
\f_1^{p_2}=\f_2^{p_1} 
\qquad \text{with}\ 1<p_1<p_2\ \text{and}\ \gcd(p_1,p_2)=1.
\end{align}
The exponents are positive since $\D(\f_i)>0$; they are greater than 1 since otherwise one of the $\f_i$ can be eliminated in favor of the other; they are coprime for irreducibility; and the ordering is a matter of convention.  This implies that the dimensions of the fields are related by $\D(\f_i) = p_i \cdot \d$ for some positive $\d$.  The unitarity condition \eqref{unitarity} implies $\d>1/p_1$.  
The inequality is strict, since if $\d=1/p_1$ then $\f_1$ has dimension 1, and so is free, so should not satisfy a nontrivial chiral ring relation such as \eqref{CRex2}. 

The uniformizing parameter for this cone is $u = \vev{\f_1}^{1/p_1} = \vev{\f_2}^{1/p_2}$ which has dimension $\D(u) = \d > 1/p_1$.  Note that even though $\D(\f_i) >1$, if $\D(\f_1) < p_1$ then $\D(u)<1$.  This does not contradict unitarity since $u$ is not the vev of any field in the SCFT.  The allowed spectrum of $\d$ can then cover the whole list of scale invariant special K\"ahler geometries appearing in the first eight rows of table \ref{table:eps}, not just the regular ones with $m=0$.

Note that the uniformizing parameter can always be written as a rational monomial in the $\vev{\f_i}$, $u \sim \vev{\f_1}^{a_1} \vev{\f_2}^{a_2}$ for integers $a_i$ such that $a_1 p_1 + a_2 p_2=1$ (which exist since $\gcd(p_1,p_2)=1$).

\paragraph{Example 3: Multi-component hypersurface CB.}  

A broader class of chiral rings are those generated again by two fields but which are not irreducible.  The ideal is principle, $\cI^0=\vev{P^0}$, with generating relation of the form
\begin{align}\label{CRex3}
P^0 &:= (\f_1)^{n_0} (\f_2)^{n_\infty} 
\prod_{a=1}^N (\f_1^{p_2}- \w_a \f_2^{p_1})^{n_a}
& \text{where}\ 
\begin{cases}
1 \le p_1 \le p_2, \ 
\gcd(p_1,p_2)=1, \\
0\le n_0,\ 0\le n_\infty,\ 0 < n_a \in\Z,\\
\w_a\in\C^* \
\text{and}\ \w_a\neq\w_b\ \text{for}\ a\neq b.
\end{cases}
\end{align} 
The CB is then the union of the sets
\begin{align}\label{CRex3a}
V(\cI^0) = \{\f_1=0\}^{\text{sgn}(n_0)} 
 \bigcup  \{\f_2=0\}^{\text{sgn}(n_\infty)} 
\bigcup_{a=1}^N \{\f_1 = \w_a \f_2\} ,
\end{align}
where for any set $S$, 
\begin{align}\label{}
S^{\text{sgn}(n)}:=
\begin{cases}
\varnothing & \text{if $n=0$},\\
S & \text{if $n>0$}.
\end{cases}
\end{align}
There are a few things to note in comparison to the previous example.  First, if any of the $n_a$ is greater than one, there is a corresponding multiplicity in the defining ideal which is not reflected in the geometry.  Second, each factor defines an equivalent conical geometry, $\cC_a := \{\f_1 = \w_a \f_2\}$, so the ratios of the $\w_a$ (which are invariants of the chiral ring) also do not have a reflection in the CB geometry.  Third, there are two special values of $\w_a$, namely $\w=0$ and $\w=\infty$, for which the corresponding cones, $\cC_0 := \{\f_1=0\}$ and $\cC_\infty := \{\f_2=0\}$, have uniformizing parameters $u_0 := \vev{\f_1}$ and $u_\infty:=\vev{\f_2}$, respectively, which, typically, have different dimensions from the uniformizing parameter, $u:= \vev{\f_1}^{1/p_1} = \vev{\f_2}^{1/p_2}$ of the other cones.  Thus, the CB described by \eqref{CRex3} looks like a bouquet of $N+\text{sgn}(n_0)+\text{sgn}(n_\infty)$ cones, all identical except for at most two, depending on whether $n_0>0$ or $n_\infty>0$.  Finally, we have chosen $\gcd(p_1,p_2)=1$ to make the components and their multiplicities, $n_a$, explicit; however, unlike the previous example, we now allow $1=p_1=p_2$ for $N+\text{sgn}(n_0)+\text{sgn}(n_\infty)>1$ since this gives a non-trivial bouquet of many cones (all with the same geometry).

\paragraph{Example 4: A non-complete intersection CB.}  

Another simple way to get a bouquet of $N$ cones is in a ring with $N$ generating fields modulo an ideal generated by $N(N-1)/2$ relations,
\begin{align}\label{CRex4}
\cI^0 = \vev{\, \f_i  \f_j\, ,\, 1\le i <j\le N\,},
\end{align}
which gives a CB as a union of the sets
\begin{align}\label{CRex3a}
V(\cI^0) = \bigcup_{i=1}^N \{\f_j = 0\ \text{for all}\ j\neq i\}.
\end{align}
This is thus a bouquet of $N$ cones each with its own uniformizing parameter given by $u_i = \vev{\f_i}$, and each cone can be any one of the eight regular $m=0$ scale-invariant cones from table \ref{table:eps}.  Note that for $N\ge3$ this chiral ring is not a complete intersection, so is not equivalent to the chiral ring of example 3.

\paragraph{Example 5: A hybrid of examples 3 and 4.}

One can combine examples 3 and 4 to get a class of chiral rings described by $M$ pairs of generating fields $\f_i^I$, $i=1,2$ and $I=1,2,\ldots,M$, with $M(M-1)/2$ relations of the form $0=P^0_I P^0_J$ for $I<J$ where each $P^0_I$ is of the form \eqref{CRex3}.  This then describes a bouquet of $\sum_I N_I(N_I-1)/2$ cones, and each component cone can be any scale-invariant entry (including those with $m\neq0$) in table \eqref{table:eps}.

\paragraph{Example 6: Another non-complete intersection CB.}

One can extend example 2 to arbitrarily many generating fields, $\{\f_p\}$ with index $p$ running over some set.  Take them to all have distinct commensurate dimensions, and label them by the integers $\D(\f_p) = p \d$, $p\in\N$, with $\gcd(\{p\})=1$.  Then the ideal 
\begin{align}\label{}
\cI^0 = \vev{\,{\f_p}^q-{\f_q}^p\,,\, \text{all}\ p\neq q\,}
\end{align}
describes a single cone with uniformizing parameter $u = {\vev{\f_p}}^{1/p}$ of dimension $\D(u)=\d$.  The condition that $\gcd(\{p\})=1$ ensures that this is an irreducible algebraic variety.

\paragraph{Example 7: An infinitely-generated CB chiral ring.}

This is the same as the last example but with the index set $\{p\}$ taken to be infinite, for example, let it be the set of all primes.  This shows that it is also possible for the chiral ring to not be finitely generated.

\paragraph{Example 8: Non-reduced CB chiral rings.}

We can add any number of nilpotent generators, $\psi_k$ satisfying ${\psi_k}^{\ell_k}=0$, to any of the above examples without changing the geometry of the CB.  Furthermore, we can add arbitrary ``extra" generators to the ideal as long as they have a nilpotent element in every term.  (I.e., we can add any generator to $\cI_0$ which is in the nilradical of $\cI_0$ without changing $V(\cI_0)$.)  A simple example is $R^0_\text{CB}=\C[\f_1,\f_2,\psi]/\cI^0$ with
\begin{align}\label{}
\cI^0 = \vev{\ \f_1^2 - \f_2^3\ ,
\ \psi^4\ ,
\ \psi \f_1 - \psi^2 \f_2\ },
\end{align}
which describes a single cone with uniformizing parameter $u = \vev{\f_1}^{1/3} = \vev{\f_2}^{1/2}$.

\section{SW curve and form for SCFT CBs}
\label{appD}

We show here via a few examples from appendix \ref{appA} how to write SW curves and forms encoding the special K\"ahler structure of rank-1 scale-invariant CBs.

Since, by the rank-1 assumption \eqref{CBrk1}, the scale-invariant CB is a bouquet of cones, $\cCb^0 = \bigvee_a \cCb_a$, with each cone $\cCb_a$ one of the entries in the first eight rows of table \ref{table:eps}, we can read off the corresponding SW curves for each $\cCb_a$ cone from the corresponding entry in table \ref{kodaira}.  In these curves and form, $u$ should be $u_a$, the uniformizing parameter of $\cCb_a$.  Each $u_a$ can be written (though not uniquely) as a rational function in the $\f_i$ which enter the chiral ring.  Furthermore, it is not hard to write a single expression $u=R(\f_i)$ with $R$ a rational function so that $u\sim u_a$ is a uniformizing parameter for each $\cCb_a$ component.  Denote by ${\bf m} = \{ m_a\}$ a choice of integer $m_a$ for each $\cCb_a$ component.  Then, more generally, there is a rational $u_{\bf m}=R_{\bf m}(\f_i)$ such that $u_{\bf m} \sim u_a^{m_a}$ on each component.  For example, here are some expressions for $u_{\bf m}=R_{\bf m}(\f_i)$ for two classes of examples from appendix \ref{appA}:

\paragraph{Example 3: Multi-component hypersurface CB.} 

Here $\cCb^0 = V(\cI^0)$ with $\cI^0=\vev{P^0}$ as in \eqref{CRex3} where we write
\begin{align}\label{}
P^0=P_0 P_\infty \prod_{a=1}^N P_a
\end{align}
with $P_0 :=(\f_1)^{n_0}$, $P_\infty := (\f_2)^{n_\infty}$, $P_a :=(\f_1^{p_2}- \w_a \f_2^{p_1})^{n_a}$.  Then $\D(\f_1)=p_1\d$ and $\D(\f_2)=p_2\d$, so $\D(P_0)=n_0 p_1 \d$, $\D(P_\infty)=n_\infty p_2 \d$, and $\D(P_a)=n_a p_1 p_2\d$.   $\cCb^0$ has $N+2$ cones: one with uniformizing parameter $u_0=\f_1$ when $P_0=0$, one with $u_\infty=\f_2$ when $P_\infty=0$, and $N$ with $u_a=\f_2^s/\f_1^t$ when $P_a = 0$.  Here $s$ and $t$ are positive integers such that $s p_2 - t p_1 = 1$, which exist since $\gcd(p_1,p_2)=1$.  Then we can write, e.g., 
\begin{align}\label{ex3uu}
u_{\bf m} = 
(u_0)^{m_0} M_0  + 
(u_\infty)^{m_\infty} M_\infty +
\sum_{a=1}^N
(u_a)^{m_a} M_a 
\end{align}
where
\begin{align}\label{}
M_0 &:= \frac{(P_\infty \prod_b P_b)^{n_0 p_1}}{(P_0)^{n_\infty p_2+\sum_b n_b p_1 p_2}}, \nonumber\\
M_\infty &:= \frac{(P_0 \prod_b P_b)^{n_\infty p_2}}{(P_\infty)^{n_0 p_1+\sum_b n_b p_1p_2}}, \\
M_a &:= \frac{(P_0 P_\infty \prod_{b\neq a} P_b)^{n_a p_1 p_2}}{(P_a)^{n_0 p_1+n_\infty p_2+\sum_{b\neq a} n_b p_1 p_2}}.
\nonumber
\end{align}
Then $u_{\bf m}$ is a single rational function of the $\f_i$ which gives an integer $m_a$ power of the uniformizer for each conical component.  In fact, there is great freedom in choosing $u_{\bf m}$, since any $M_a$ in \eqref{ex3uu} could just a well be replaced by $L_a M_a$ where $L_a$ is any dimensionless rational function of the $\f_i$ (not vanishing on the $\cC_a$ component).

\paragraph{Example 4: A non-complete-intersection CB.} 

Here, from \eqref{CRex4}, $\cI^0=\vev{P_{ij}}$ with $P_{ij}=\f_i\f_j$ has $N$ cones, one each with $u_i=\f_i$ when $\f_j=0$ for all $j\neq i$.  We can simply write, e.g., 
\begin{align}\label{}
u_{\bf m} = \sum_i (\f_i)^{m_i}
\end{align}
for a universal uniformizer.

\bigskip

With such a collection of uniformizers, $u_{\bf m}=R_{\bf m}(\f_i)$, giving arbitrary powers of uniformizing parameters on each component, we can then write a single SW curve for the whole SCFT CB, $\cCb^0$, by choosing its $f$ and $g$ coefficients (in regular gauge) as simply $f = u_{\bf m}$ and $g = u_{\bf m'}$ for an appropriate choice of $\bf m$ and $\bf m'$ with entries from the list given in table \ref{kodaira}.  Furthermore, in regular gauge, the 2-form is then $\Omega = du_{\bf 1} \wedge dx/y$ where ${\bf 1} = \{ 1, 1, \ldots, 1\}$.  This shows that there is an infinite ambiguity in the choice of SW curve used to describe any given SCFT $\cCb^0$ geometry.


\section{Examples of deformed rank-1 CB chiral rings}
\label{appB}

Here we explore the effect of relevant and marginal deformations of the SCFT on the description of the CB as a complex variety.  

Let's suppose that the SCFT chiral ring is finitely generated, so that the CB chiral ring of the undeformed SCFT is $R^0_\text{CB} = \C[\f_1,\ldots,\f_v]/\cI^0$ with $\cI^0 = \vev{P^0_1, \ldots, P^0_r}$, where the $r$ generating relations, $P^0_k=0$, are weighted homogeneous polynomials in the $v$ variables, $\f_i$, with positive weights $D(\f_i) := \d_i$.  Then, by assumption \eqref{CBass}, the CB is the complex algebraic variety $\cCb^0 = V(\cI^0)$.  (Recall that the ``$0$" superscripts denote undeformed quantities.)

Since the deformation does not change the chiral field content of the SCFT, we expect that the CB chiral ring of the deformed theory will be of the form $R_\text{CB} = \C\{\f_1,\ldots,\f_v\}/\cI$ for some deformed ideal $\cI$.  A general analytic deformation of the ideal may be complicated; we sidestep this by assuming that a set of generators of the deformed ideal are given by polynomial deformations of the generators of the undeformed ideal.  That is, upon deformation, assume further that the CB chiral ring is 
\begin{align}\label{RCBdef1}
R_\text{CB} = \C[\f_1,\ldots,\f_v]/\cI
\qquad\text{with}\quad
\cI = \vev{P_1,\ldots,P_r}
\quad\text{where} \quad P_k = P^0_k + \d P_k,
\end{align}
such that each deformed relation, $P_k$, is a polynomial in the $\f_i$ of highest weighted degree the same as that of $P^0_k$.  The latter part of this assumption ensures that the deformation parameters all have non-negative dimension, so correspond only to relevant or marginal deformations of the SCFT.  

Not all deformations satisfying the last assumption are compatible with $\cN=2$ supersymmetry.  For instance, $\cN=2$ supersymmetry-preserving deformations do not lift any flat directions.  This may impose additional constraints on the allowed deformations in cases where the CB chiral ring has a syzygy --- i.e., when there are relations among the generators of $\cI^0$, so $V(\cI^0)$ is not a complete intersection.  This, together with \eqref{RCBdef1}, imply that the deformed CB, $\cCb = V(\cI)$, is asymptotically the same as $\cCb^0$ in the limit of large vevs, $\vev{\f_i} \gg 1$, which is another condition that relevant and marginal $\cN=2$ deformations must satisfy \cite{paper1}.  

In particular, the condition that none of the asymptotic flat directions are lifted upon deformation is a necessary condition for a \emph{flat deformation}.  More generally, flatness seems to capture the notion that the deformation is continuous, so it is natural to also make the assumption that the general CB chiral ring is a flat deformation of the SCFT CB chiral ring.  See, e.g., \cite{GLS07} for the definition of flatness.  Flat deformations have the property that any such deformation can be described as a special case of a ``miniversal" deformation \cite{GLS07}.

It is not clear to us whether the assumptions of equi-dimensional deformations given by polynomial deformations of the generators of the SCFT chiral ring ideal as in \eqref{RCBdef1} are enough to imply the flatness assumption.  Note that although flat deformations also necessarily have the form \eqref{RCBdef1}, the condition on the degree of the generators on the deformed ideal is not implied by flatness.  In field theory terms, flat deformations may allow RG irrelevant as well as relevant and marginal deformations.  

We now make this discussion more concrete by computing the (flat) miniversal deformations in some examples.   The numbers refer to the examples of rank-1 SCFT chiral rings given in appendix \ref{appA}, but the deformations are only computed for specific (simple) cases of those examples.

\paragraph{Example 2.}  Consider a SCFT with CB given by a single cone $\cCb^0 = V(\cI^0)$ described by $\cI^0=\vev{P_1^0}$ with
\begin{align}\label{}
P^0=\f_3^2 + \f_2^3,
\end{align}
and $D(\f_k)=k\d$.  The uniformizing parameter for this cone is $u := \f_k^{1/k}$ with $D(u) = \d$.   Then its possible deformations (modulo redefinitions of the $\f_k$ of the form $\f_k \to \m_k \f_k + \n_k$ for suitable complex parameters $\m_k$ and $\n_k$) are $\cI = \vev{P}$ with
\begin{align}\label{}
P = P^0 + M_4 \f_2 + M_6
\end{align}
where the deformation parameters have dimensions $D(M_a) = a\d$, so are both relevant.   In this case the deformed CB is described by the plane curve $\cCb = V(\cI)$ which is easily seen to describe a genus-1 Riemann surface with a single puncture at $(\f_2,\f_3)=(\infty,\infty)$.

\paragraph{Example 3.}  Consider the bouquet of three cones given by $\cI^0 = \vev{\,  P^0 \,}$ with
\begin{align}\label{}
P^0 := \f_1 (\f_1-\f_2)^2 \f_2.
\end{align}
Note that the $\f_1=\f_2$ cone has multiplicity 2.
This is homogeneous for the $\f_j$ having the same dimensions, $\D(\f_1)=\D(\f_2)=\d$.
Its miniversal deformation is given by $\cI=\vev{\, P\,}$ with
\begin{align}\label{}
P &= P^0 + m_0 \f_1(\f_1-\f_2)\f_2^2 + m_1 \f_2^3 + m'_1 \f_1\f_2^2
+ m_2 \f_2^2 + m'_2 \f_1 \f_2 + m''_2 \f_1^2
\nonumber\\
&\qquad\ \,\text{}+ m_3 \f_2 + m'_3 \f_1 + m_4,
\end{align}
where the nine $m_j$ deformation parameters have dimensions $j\cdot \d$, so are all relevant except for $m_0$ which is marginal.  The resulting generically deformed CB is a genus 3 Riemann surface with four punctures.  Note that the marginal deformation has separated the single multiplicity-2 $\f_1=\f_2$ cone to two (multiplicity-1 asymptotic) cones at $\f_1=\f_2$ and $\f_1=(1+m_0)\f_2$.   

Note that linear redefinitions of $\f_1$ and $\f_2$ can be used to put any 3 distinct cones at, say, $\f_1=0$, $\f_2=0$, and $\f_1=\f_2$, so it is only the relative position of fourth and higher cones that correspond to deformation parameters.  In particular, this  makes clear some behaviors of flat deformations which suggests that flatness may not be the correct criterion for physically admissable deformations.   First, if there are three or fewer cones (counting with multiplicity), then there is no flat marginal deformation parameters which separates multiplicity-2 cones as above.   Second, if there are five or more cones (counting with multiplicity) then there will be irrelevant flat deformation parameters.

\paragraph{Example 4.}  Consider the bouquet of three cones given by the ideal in $\C[\f_1,\f_2,\f_3]$
\begin{align}\label{}
\cI^0 = \vev{\,  P_{12}^0 \, ,\, P_{23}^0 \, ,\, P^0_{31} \,}
\qquad \text{with}\quad 
P_{ij}^0 :=\f_i\f_j .
\end{align}
The generators satisfy further relations (a ``syzygy") $\f_1 P^0_{23} = \f_2 P^0_{31} = \f_3 P^0_{12}$, which ensure that the three $P^0_{ij}=0$ equations in three variables still have a 1-dimensional solution set.  Its miniversal deformation is given by the deformed ideal 
\begin{align}\label{}
\cI=\vev{\,  P_{12} \, ,\, P_{23} \, ,\, P_{31} \,}
\qquad\text{with}\quad
P_{ij} = P_{ij}^0 + M_j \f_i + M_i M_j ,
\end{align}
where the $M_j$ are three complex deformation parameters.  
(This can be computed using, e.g., a computer algebra system like {\sc Singular}.)  It is clear from the grading of the generators that $M_j$ must have the same dimension as $\f_j$, so all three are relevant deformations.  Note that none of the $M_j$ can be reabsorbed in a shift of the $\f_j$.  The syzygy is deformed so that for generic values of the $M_j$ any one of the $P_{ij}$s is implied by the other two.  This implies that the resulting generically deformed CB is a genus zero Riemann surface with 3 punctures.

\paragraph{Example 6.}  Consider the single cone given by 
\begin{align}\label{}
\cI^0 = \vev{\,  P_{23}^0 \, ,\, P_{25}^0 \, ,\, P^0_{35} \,}
\qquad \text{with} \quad 
P^0_{pq} := {\f_p}^q - {\f_q}^p
\end{align}
with $p,q \in \{2,3,5\}$ so $\D(\f_p) = p\d$ and the uniformizing parameter of the cone has dimension $\D(u)=\d$.  Its relevant miniversal deformations are given by $\cI = \vev{\,  P_{23} \, ,\, P_{25} \, ,\, P_{35} \,}$ with
\begin{align}\label{}
P_{23} &= P^0_{23} + Q, \nonumber\\
P_{25} &= P^0_{25} + {\f_2}^2 \cdot Q + M_1 (\f_5 - \f_2 \f_3),\\
P_{35} &= P^0_{35} - ({\f_3}^3 -{\f_2}^2 \f_5 +{\f_2}^3\f_3) \cdot Q + (\f_5 - \f_2 \f_3) \cdot  R, \nonumber
\end{align}
where
\begin{align}\label{}
Q &:= M''_1 \f_5 + M'''_1 \f_2 \f_3 + M'_2 {\f_2}^2 + M'_3 \f_3
+ M'_4 \f_2 + M'_6,
\nonumber\\ 
R &:= M_1 \f_5 + M'_1 {\f_2}^3\f_3 + M_2 {\f_2}^4 
+ M_3 {\f_2}^3 \f_3 + M_4 {\f_2}^3 + M_5 \f_2 \f_3
\\
& \qquad + M_6 {\f_2}^2 + M_7 \f_3 + M_8 \f_2 + M_{10}. \nonumber
\end{align}
The sixteen relevant $M_j$ have dimensions $j\d$.  (There are two other flat deformations which are irrelevant with dimensions $-\d$.)  The resulting generically deformed CB turns out to be the union of a genus 1 Riemann surface with one puncture together with ten isolated (multiplicity-1) points.  Thus this deformed CB does not have uniform dimension, but components of both dimension 0 and 1.  This is another illustration of the point made in section \ref{s:dis} that if SCFTs with non-trivial chiral ring relations are allowed, then the existence of rank-0 IR fixed points is a natural consequence.  Note, however, that in this case there is no argument that the rank-0 fixed points need be SCFTs:  they may be trivial or IR free.

\end{appendix}

\bibliographystyle{JHEP}
\providecommand{\href}[2]{#2}\begingroup\raggedright\endgroup
\end{document}